\documentclass[a4paper,11pt]{article}
\pdfoutput=1 

\usepackage{jcappub} 

\usepackage[T1]{fontenc} 
\usepackage[utf8]{inputenc}
\usepackage{hyperref}
\usepackage{cprotect}
\usepackage{float}
\usepackage{subcaption}
\usepackage{caption}
\usepackage{times}

\title{{\sc gramses}: a new route to general relativistic $N$-body simulations in cosmology. Part II.~Initial conditions}

\author[a]{Cristian Barrera-Hinojosa}
\author[a]{and Baojiu Li}

\affiliation[a]{Institute for Computational Cosmology, Department of Physics, Durham University,\\ Durham DH1 3LE, UK}

\emailAdd{cristian.g.barrera@durham.ac.uk}
\emailAdd{baojiu.li@durham.ac.uk}

\abstract{We address the generation of initial conditions (ICs) for {\sc gramses}, a code for nonlinear general relativistic (GR) $N$-body cosmological simulations recently introduced in Ref.~\cite{Barrera-Hinojosa:2019mzo}. {\sc gramses} adopts a constant mean curvature slicing with a minimal distortion gauge, where the linear growth rate is scale-dependent, and the standard method for realising initial particle data is not straightforwardly applicable.
A new method is introduced, in which the initial positions of particles are generated from the displacement field realised for a matter power spectrum as usual, but the velocity is calculated by finite-differencing the displacement fields around the initial redshift. In this way, all the information required for setting up the initial conditions is drawn from three consecutive input matter power spectra, and additional assumptions such as scale-independence of the linear growth factor and growth rate are not needed. We implement this method in a modified {\sc 2LPTic} code, and demonstrate that in a Newtonian setting it can reproduce the velocity field given by the default {\sc 2LPTic} code with subpercent accuracy. We also show that the matter and velocity power spectra of the initial particle data generated for {\sc gramses} simulations using this method agree very well with the linear-theory predictions in the particular gauge used by {\sc gramses}. Finally, we discuss corrections to the finite difference calculation of the velocity when radiation is present, as well as additional corrections implemented in {\sc gramses} to ensure consistency. This method can be applied in ICs generation for GR simulations in generic gauges, and simulations of cosmological models with scale-dependent linear growth rate.}

\begin{document}
\maketitle
\flushbottom

\section{Introduction}\label{sec:introduction}

Recently we have introduced a new code for general relativistic $N$-body simulations, {\sc gramses} \cite{Barrera-Hinojosa:2019mzo}. In this paper, we address the issue of generating particle initial conditions (ICs) -- namely, positions and velocities -- for simulations using this and other general relativistic codes. The proper generation of ICs is an essential part of the pipeline since the gauge choice in {\sc gramses} -- a combination of constant mean curvature (CMC) slicing and minimal distortions (MD) -- means that standard ICs generated for Newtonian $N$-body simulations are not in the gauge used by {\sc gramses}. 
This is because, 
in the linear regime, where the synchronous gauge is well defined, the density perturbation and velocity field used in Newtonian simulations are equal to the corresponding quantities in the synchronous and Newtonian gauges, respectively~\cite{Chisari:2011iq,Flender:2012-UE,Hwang:2012}. 
Standard initial condition codes, such as 2{\sc lpt}ic \cite{Scoccimarro:1997gr,Crocce2006:2LPT}, {\sc grafic}~\cite{Bertschinger:2001ng} or {\sc mpgrafic}~\cite{Prunet:2008fv}, are tailored for this type of simulations and use parameterisations for the growth factor and growth rate that could break down in other gauges. It is therefore {necessary to modify the methods} for generating initial particle data to make them compatible with {\sc gramses} simulations. 

Contrary to numerical relativity codes based on the hyperbolic formulations of General Relativity (GR), such as Refs.~\cite{Mertens:2015ttp,Macpherson:2016ict,Daverio:2019gql}, {\sc gramses} implements the fully constrained formulation, where the gravitational sector does not require the specification of ICs as the time evolution is fully encoded in the matter sources, and therefore the initial particle data is the only type of ICs needed for such simulations, in an analogous way to standard Newtonian simulations.

For cosmological simulations of this type, the usually-used approach to generate particle positions from a given density field is to use the displacement vector field, where one relies on the fact that the initial density field is nearly homogeneous, and that tiny perturbations arise from slightly displacing the particles from a regular configuration such as a grid or glass \cite{Baugh95:glass}. At first order in perturbation theory, the solution {of the displacement} is simply given by the Zel'dovich approximation \cite{Zeldovich:1969sb}. On the other hand, the velocity can be calculated by using the first-order continuity equation, although this requires additional information about the density field, in particular its time evolution, which at linear level is encoded in the linear growth rate $f_1$. However, the latter quantity depends on both the gauge used for the overdensity, as well as the underlying theory of gravity, and one has to be careful about the parameterisation of $f_1$ adopted even in a non-relativistic scenario. For instance, it is well-known that the growth rate (and growth factor) becomes scale-dependent in many modified gravity or dark energy models~\cite{Linder,Narikawa:2009ux}. Another potential limitation of this method is that, {as we shall note later,} it is possible that the overdensity and velocity fields from a given gauge do not satisfy the `standard' (or `Newtonian') form of the continuity equation even if the conservation of the energy momentum tensor holds, since in general {the continuity} equation also contains curvature perturbation terms, which for Newtonian $N$-body simulations can be properly taken into account \cite{Fidler:2015npa}. A general approach to tackle this problem has been implemented in {\sc falconIC} code \cite{Valkenburg:2015dsa}, which is capable of generating ICs for a wide range of theories of gravity, including those having perturbations in non-standard matter components at high redshift, as well as for models with imperfect fluids such as neutrinos.

In order to circumvent the gauge issues when calculating the velocity field for {\sc gramses} simulations, we propose a finite difference method in which, roughly speaking, two subsequent particle {snapshots close to the initial redshift, generated using the same random number seeds,} are compared to obtain the velocity field connecting them. This is a very straightforward but versatile approach due to the advantage of being independent of an explicit parameterisation of the growth factor, as the information about the evolution of the density field is drawn from the comparison of the snapshots. We have implemented this finite difference method by modifying the 2{\sc lpt}ic code. As the input for solving the displacement vector problem at a given redshift, the default 2{\sc lpt}ic code uses a matter power spectrum rather than the density field itself. In order to calculate the spectrum in the CMC gauge, we use a modified version of the {\sc camb}~\citep{CAMB} which implements suitable gauge transformations from the default synchronous gauge used in such code. Since for the finite difference calculation of the velocity we need the displacement vectors from two subsequent snapshots, in the modified 2{\sc lpt}ic code we additionally input two neighbouring matter power spectra (one at a slightly higher redshift while the other at a slightly lower redshift than the {true initial time}). Then, the code realises these power spectra to calculate the particle positions in the standard way, but the velocity is calculated by finite differencing the particle displacements from the two neighbouring snapshots. In this way, the velocity can be calculated even in cases where basic assumptions of the default 2{\sc lpt}ic code, such as scale-independent growth rate, are violated. 

The rest of this paper is organised as follows. Given that the gauge choice for {\sc gramses} and its role in the fully constrained formulation of GR have been described in details in the main paper, Ref.~\cite{Barrera-Hinojosa:2019mzo}, in Section~\ref{sec:gramses-gauge} we only briefly recall some aspects that are relevant for the current project. In Section~\ref{sec:gauge-transform} we discuss the gauge transformations that connect the CMC-MD gauge to the synchronous and Newtonian gauges, and that allow us to deal with the gauge issues behind the generation of ICs. In Section~\ref{sec:IC-displacement-method} we briefly discuss the standard displacement vector method for the generation of the particle ICs from an initial matter density field, and we point out its potential limitations when dealing with general relativistic $N$-body simulations, while in Section~\ref{sec:nbody-simulations} we explain how it is compatible with Newtonian simulations. In Section~\ref{sec:IC-gramses-gauge} we show that this method is also compatible with the generation of particle positions for {\sc gramses} by identifying carefully the overdensity variable used in the simulation \cite{Valkenburg:2015dsa}, but that velocities remain affected by the gauge dependence. As a way to overcome this gauge issue, in Section~\ref{sec:FD-method} we discuss how to calculate the initial particle velocities via a finite difference method. Then, in Section~\ref{sec:results} we present results regarding the ICs for {\sc gramses} simulations, as well as a comparison to the standard method. 
Finally, we wrap up in Section~\ref{sec:conclusion} with some conclusions and outlook.

As in the main {\sc gramses} paper~\cite{Barrera-Hinojosa:2019mzo}, throughout this paper we adopt the $(-,+,+,+)$ signature for the spacetime metric as well as the unit $c=1$. Greek indices run from 0 to 3, whereas Latin ones from 1 to 3 only, with repeated indices implying summation.

\section{The gauge choice in {\sc gramses}}
\label{sec:gramses-gauge}

The gauge choice for {\sc gramses} simulations and its implications in the constrained formulation of GR has been discussed in detail in the main paper of this series Ref.~\cite{Barrera-Hinojosa:2019mzo}, and here we only recall some relevant points for completeness. In the $3+1$ formalism the spacetime metric takes the form
\begin{equation}\label{eq:adm-metric}
{\rm d}s^2 = g_{\mu\nu}{\rm d}x^{\mu}{\rm d}x^{\nu} = -\alpha^2 {\rm d}t^2+\gamma_{ij}\left(\beta^i{\rm d}t+{\rm d}x^i\right)\left(\beta^j{\rm d}t+{\rm d}x^j\right),
\end{equation}
where $\gamma_{ij}$ is the induced metric on the spatial hypersurfaces, while the lapse function $\alpha$ and shift vector $\beta^i$ represent the diffeomorphism invariance of GR. Even though in principle we have the complete freedom to choose the gauge, in practice not all options are physically or numerically convenient. As an example, the geodesic slicing (or synchronous gauge) is characterised by $\alpha=1$ and $\beta^i=0$, but it can become ill-defined when shell crossing (or trajectory crossing) occurs, as is expected for collisionless particles in cosmological simulations. For this type of simulations, a convenient prescription for $\alpha$ is by applying the so-called Constant Mean Curvature (CMC) slicing condition~\cite{Smarr-York:MEC-1978}, in which the trace of the extrinsic curvature of the spatial hypersurfaces is fixed as a function {of time} only, 
\begin{equation}
K=-3H(t)\,,\label{eq:CMC}
\end{equation}
where $H\equiv\dot{a}/a$ represents a fiducial Hubble parameter (with $a$ being a fiducial scale factor). Here, $H$ (and $a$) is just a prescribed function for fixing the spacetime foliation and in principle does not {have to} represent average (or background) properties of the actual universe. Nevertheless, we can still fix $H$ such that it satisfies some `reference' Friedmann equations~\cite{Giblin:2017juu,Giblin:2018ndw}. Under the CMC slicing, the lapse can be found by solving
\begin{equation}\label{eq:cmc-elliptic}
\bar{\nabla}^2(\alpha\psi)=\alpha\left[2\pi\psi^{-1}(s_0+2s)+\frac{7}{8}\psi^{-7}\bar{A}_{ij}\bar{A}^{ij}+\psi^5\left(\frac{5{K}^2}{12}+{2\pi}\rho_\Lambda\right)\right]-\psi^5\dot{K}\,.
\end{equation}
Here, $\psi$ represents the conformal factor which connects 
$\gamma_{ij}$ to the conformal metric $\bar{\gamma}_{ij}$ through $\gamma_{ij}=\psi^4\bar{\gamma}_{ij}$, with $\bar{\gamma}\equiv\det(\bar{\gamma}_{ij})=1$, $\bar{\nabla}^2\equiv\bar{\gamma}^{ij}\bar{D}_i\bar{D}_j$ is the covariant Laplace operator associated with $\bar{\gamma}_{ij}$ (and $\bar{D}_i$ the associated covariant derivative), and $\bar{A}_{ij}$ is the traceless part of the extrinsic curvature tensor $\bar{K}_{ij}$. Furthermore, $\rho_\Lambda$ is the dark energy density appearing in the reference Friedmann equations, and the conformal matter source terms are defined as
\begin{align}
s_0&=\sqrt{\gamma}\rho\,,\label{eq:s_0}\\
s_i&=\sqrt{\gamma}S_i\,,\\
s_{ij}&=\sqrt{\gamma}S_{ij}\label{eq:s_ij}\,.
\end{align}
and $s=\gamma^{ij}s_{ij}$, where $\rho$, $S_i$ and $S_{ij}$ are the projections of the energy momentum tensor, $T^{\mu\nu}$, onto the spatial hypersurfaces. It can be shown that Eqs.~\eqref{eq:s_0}-\eqref{eq:s_ij} are analogous to the usual `comoving' matter source terms and correspond to those actually determined numerically such as in a Cloud-in-Cell (CIC) scheme~\cite{masaru2015:NR-book}. In particular, in Section~\ref{sec:IC-gramses-gauge} we will show that at the linear level the `density' contrast for $s_0$, defined as $\delta s_0/s_0$, corresponds to local fluctuations in the particle number density rather than the relativistic energy density $\rho$. In practice, this is more convenient for $N$-body simulations as in these we are interested in following `particles' rather than the full `energy density field' itself: a given particle can contribute different energy densities at different positions, and the relativistic correction effect can be calculated {once we have} the local values of the spatial metric $\gamma_{ij}$. This has important implications on the generation of ICs as we will see in Section~\ref{sec:IC-displacement-method}.

After adopting the CMC slicing condition Eq.~\eqref{eq:CMC}, we still have gauge freedom to choose spatial coordinates on each spatial hypersurface as represented by the three degrees of freedom in $\beta^i$. Instead of fixing $\beta^i$ `statically', such as in synchronous gauge (where it vanishes at all times), we can use this freedom to propagate the spatial coordinates from a hypersurface at $t$ to the next one at $t+\delta t$ in such a way that the `distortion' of local volume elements due to coordinate effects is minimised. For this purpose, {we apply} the Minimal Distortion (MD) gauge condition~\cite{Smarr-York:MEC-1978,Smarr1978:MDC}, in which we demand
\begin{equation}\label{eq:MD-condition}
{D}^i(\gamma^{1/3}\partial_t\bar{\gamma}_{ij})=0\,,
\end{equation}
with $D_i$ the covariant derivative associated with $\gamma_{ij}$. Using the MD condition Eq.~(\ref{eq:MD-condition}), the momentum constraint and evolution equation for $\gamma_{ij}$ combine into the following elliptic equation for the shift vector
\begin{equation}\label{eq:MDC-beta}
(\bar{\Delta}_L\beta)^i+(\bar{L}\beta)^{ij}\bar{D}_j\ln{\psi^6}=2\psi^{-6}\bar{A}^{ij}\bar{D}_j\alpha+16\pi\psi^4\alpha S^i,
\end{equation}
where $$(\bar{L}\beta)^{ij}\equiv\bar{D}^{i}\beta^{j}+\bar{D}^{j}\beta^{i}-2/3\bar{\gamma}^{ij}\bar{D}_k\beta^k$$ is a conformal Killing operator and $(\bar{\Delta}_L\beta)^i\equiv\bar{D}_j(\bar{L}\beta)^{ij}$ a conformal vector Laplacian. However, the MD gauge condition Eq.~\eqref{eq:MDC-beta} is actually simplified in the 
constrained formulation of GR adopted in {\sc gramses}, which allows to consistently neglect tensor modes in the metric and hence no evolution equations for gravity are required to be solved. Following Refs.~\cite{Bonazzola-FCF:2004,CorderoCarrion:2008-1}, in this scheme we make the approximations
\begin{equation}\label{eq:FCF-approx}
\bar{\gamma}_{ij}=\delta_{ij},\quad \bar{A}^{ij}_{\rm TT}=0\qquad\forall t\,,
\end{equation}
where $\bar{A}^{ij}_{\rm TT}$ is the transverse-traceless {(TT)} part of $\bar{A}^{ij}$. This approach follows the same spirit as the `waveless theories of gravity' developed originally by Isenberg 
\cite{Isenberg:2007zg} and later by Wilson and Mathews
in \cite{Wilson-M:1989}. We refer the reader to the main {\sc gramses} paper~\cite{Barrera-Hinojosa:2019mzo} for more discussion on this formalism as well as to~\cite{Bonazzola-FCF:2004,CorderoCarrion:2008-1,CorderoCarrion:2008nf} for its foundations and numerical applications to relativistic simulations of compact objects. {With} the conformal flatness approximation from Eq.~\eqref{eq:FCF-approx}, Eq.~\eqref{eq:MDC-beta} is simplified to
\begin{equation}\label{eq:MDC-CF}
(\bar{\Delta}_L\beta)^i=2\partial_j\left(\alpha\psi^{-6}\bar{A}^{ij}_{\rm L}\right)\,,
\end{equation}
where $\bar{A}^{ij}_{\rm L}$ is the longitudinal part of $\bar{A}^{ij}$. We remark that with this gauge condition, the shift vector $\beta^i$ has both scalar (longitudinal) and vector (transverse) modes, which makes it different from the shift vector appearing in the commonly-used Poisson gauge where it contains purely vector perturbations. Throughout this paper, we refer to the combinations of both CMC slicing and MD conditions as the `CMC-MD' gauge, though `CMC' will often be used for the same meaning in order to avoid cluttered notation.

\section{Gauge transformations}
\label{sec:gauge-transform}

Since the gauge issue plays an important role in the generation of ICs, in this section we shall briefly discuss gauge transformations to understand how the main quantities from the CMC-MD gauge are connected to those in the synchronous and Newtonian gauges at first order in perturbation theory. A comparison of the latter two gauges to the CMC gauge is given in Ref.~\cite{Flender:2012-UE} where this is referred to as the `Uniform Expansion' gauge, but here we will also give details on the MD gauge for the choice of spatial coordinates as well as on the equations in the 3+1 formalism. For the synchronous gauge and Newtonian gauge quantities we will stick to the convention of notation of Ref.~\cite{MB95}.

\subsection{The geometric sector}

Under an infinitesimal reparameterisation of coordinates, $x^\mu\to {x}'^\mu=x^\mu+\xi^\mu$, the metric components transform as
\begin{align}
g_{\alpha\beta}(x)={g}'_{\alpha\beta}(x)+{g}_{\mu\alpha}\partial_\beta\xi^\mu+{g}_{\mu\beta}\partial_\alpha\xi^\mu+\xi^\mu\partial_\mu g_{\alpha\beta}\,,\label{eq:metric-gauge-t}
\end{align}
where ${g}'_{\alpha\beta}$ is the spacetime metric in the new coordinate system, and we have expanded this around the original spacetime point $x^\mu$ as $${g}'_{\alpha\beta}({x}')\approx{g}'_{\alpha\beta}(x)+\xi^\mu\partial_\mu g_{\alpha\beta}(x).$$ To connect with standard perturbation theory and the different gauges used in cosmology, we linearise the $3+1$ metric in Eq.~\eqref{eq:adm-metric} around a Friedmann-Lema\^itre-Robertson-Walker (FLRW) background {with metric $\text{diag}(-1,a^2\delta_{ij})$}, which coincides with the `fiducial' background introduced through the CMC slicing condition. Then, we apply Eq.~\eqref{eq:metric-gauge-t} to obtain the transformation laws for the perturbed $3+1$ metric in terms of cosmic time $t$ and comoving spatial coordinates $x^i$. 
The conformal factor defined through the relation $\gamma_{ij}=\psi^4\bar\gamma_{ij}$ is perturbed at first order as $\psi=a^{1/2}(1-\Psi/2)$, while the perturbed conformal metric is $\bar{\gamma}_{ij}=\delta_{ij}+h_{ij}$. Therefore, the metric components of the linearised Eq.~\eqref{eq:adm-metric} are 
\begin{eqnarray}
g_{00}&=&-\left(1+2\Phi\right)\,,\label{eq:cmc_metric}\\
g_{0i}&=&\beta_i\,,\\
\gamma_{ij}&=&a^2[(1-2\Psi)\delta_{ij}+h_{ij}]\,,\label{eq:cmc_metric3}
\end{eqnarray}
where we have introduced the lapse perturbation $\Phi\equiv\alpha-1$ and $h_{ij}$ is a traceless tensor, i.e. ${\gamma}^{ij}h_{ij}=0$. Then, applying the transformation law Eq.~(\ref{eq:metric-gauge-t}) we find that the metric perturbations transform as
\begin{align}
\Phi' &= \Phi +\dot{\xi}_0,\label{eq:alpha-gauge-transform}\\
\beta_i' &= \beta_i - \dot{\xi}_i - \partial_i\xi_0 + 2H\xi_i,\label{eq:beta-gauge-transform}\\
\gamma'_{ij} &= \gamma_{ij} - \partial_i\xi_j - \partial_j\xi_i + 2a\dot{a}\delta_{ij}\xi_0,\label{eq:gamma-gauge-transform}
\end{align}
where $H\equiv\dot{a}/a=-K/3$ is the Hubble parameter fixed by the CMC foliation, Eq.~\eqref{eq:CMC}. From the trace and traceless parts of (\ref{eq:gamma-gauge-transform}) we find, respectively,
\begin{align}
   \Psi'&=\Psi-H\xi_0+\frac{1}{3}a^{-2}\delta^{ij}\partial_j\xi_{i}\,,\label{eq:cmc_psi}\\
    h'_{ij}&=h_{ij}-a^{-2}(\partial_j\xi_{i}+\partial_i\xi_{j})+\frac{2}{3}a^{-2}\delta^{kl}\partial_k\xi_{l}\delta_{ij}\,.\label{eq:h_ij-gauge-transform}
\end{align}
At the linear level, the MD condition Eq.~\eqref{eq:MD-condition} reduces to
\begin{equation}\label{eq:MD-linear-gauge-condition}
    \partial^ih'_{ij}=0\,,
\end{equation}
so that Eq.~\eqref{eq:h_ij-gauge-transform} can be used to connect the spatial components of the gauge transformation variable $\xi^\mu$ with $\partial^ih_{ij}$, which can in turn be used to link the spatial coordinates in the two gauges. It is useful to note that if $h_{ij}=0$ in a given gauge, such as in the case of Newtonian gauge (or Poisson gauge in the absence of tensor perturbations), then $\xi_i=0$ and the spatial coordinates in such a gauge are equivalent to those in the MD gauge (at first order). 

\subsection{The matter sector}
\label{sec:gauge-matter-sector}

For the matter sector, let us consider the energy-momentum tensor of the form
\begin{equation}\label{eq:energy-momentum-tensor}
    T^\mu_{\nu}=(\rho + P)u^\mu u_\nu +P\delta^{\mu}_\nu+\Sigma^\mu_\nu\,,
\end{equation}
in which $u^\mu={\rm d}x^\mu/{\rm d}\tau$ is the 4-velocity of the fluid, $\rho$ is the energy density, $P$ the pressure and $\Sigma^\mu_\nu$ the anisotropic stress tensor. Under the infinitesimal coordinate reparameterisation, the transformation law for Eq.~\eqref{eq:energy-momentum-tensor} is
\begin{equation}
    T^\alpha_{\ \beta}(x) = T'^{\alpha}_{\ \ \beta}(x) + T'^\alpha_{\ \ \nu}\partial_\beta\xi^\nu - T'^{\mu}_{\ \ \beta}\partial_\mu\xi^\alpha + \xi^\lambda\partial_\lambda T'^\alpha_{\ \ \beta}.\label{eq:T-gauge-transnf}  
\end{equation}
Using that $-T^0_{\ 0}=\rho=\bar{\rho}(1+\delta)$, where {the overbar means that} $\bar{\rho}$ is the background density {(overbar always has this meaning when applied to matter quantities throughout this paper)} and $\delta$ is the density contrast, we find that the latter transforms as
\begin{equation}\label{eq:delta-gauge-transf}
    \delta = \delta'+ 3H(1+w)\xi_0\,,
\end{equation}
where we have used the background continuity equation $\dot{\bar{\rho}}+3H\bar{\rho}(1+w)=0$, and $w\equiv\bar{P}/\bar{\rho}$ is the equation-of-state parameter for a given specie, {and again an overbar is used in $\bar{P}$ to highlight that this is the mean pressure}. Similarly, since $T^0_i=(\bar{\rho}+\bar{P})u^0u_i$, where $u^0=1$ to first order, and $T^i_j=(\bar{P}+\delta P)\delta^i_j+\Sigma^i_j$, with $\Sigma^i_i=0$, we find that the lower-index velocity transforms as
\begin{equation}\label{eq:ulowi-gauge-transf}
    u'_i=u_i-\partial_i\xi_{0}\,,
\end{equation}
while the upper-index velocity transforms as
\begin{equation}\label{eq:uupperi-gauge-transf}
    u'^i = u^i + a^{-2}\delta^{ij}\dot{\xi}_j - 2a^{-2}H\delta^{ij}\xi_j\,.
\end{equation}
Equations \eqref{eq:delta-gauge-transf}, \eqref{eq:ulowi-gauge-transf} and \eqref{eq:uupperi-gauge-transf} can be used to transform the density and velocity from the CMC-MD gauge to other gauges. In particular, by using the aforementioned fact that $\xi_i=0$ when connecting the MD gauge with Newtonian gauge in Eq.~\eqref{eq:uupperi-gauge-transf} we find that the 4-velocity $u^i$ is actually the same in both gauges. However, this is only true for the upper-index velocity $u^i$, while $u_i$ transforms with $\partial_i\xi_0$ as shown by Eq.~\eqref{eq:ulowi-gauge-transf}. We will come back to this point later.

For the sake of completeness, in Appendix~\ref{appendix:mapping} we include additional details about the mapping between the linearised fields and evolution equations in the CMC-MD gauge and their counterparts in the synchronous and Newtonian gauges.

\section{The generation of initial conditions}

Before discussing the method for the generation of ICs for particles, we remark that in the constrained formulation implemented in {\sc gramses}  \cite{Bonazzola-FCF:2004,CorderoCarrion:2008-1}, the initial data for the metric is entirely determined by the initial particle data, as there are no dynamical degrees of freedom in the metric (tensor modes) due to the approximations in Eq.~\eqref{eq:FCF-approx}.

At early times, when fluctuations around the FLRW background universe are small, it is usually assumed that standard perturbation theory is accurate and allows to set ICs for $N$-body simulations, which then take care of the nonlinear evolution throughout the late-time universe. This is usually done by solving the linear perturbation equations for the coupled cosmic fluid numerically in a Boltzmann code such as {\sc camb}~\cite{CAMB} or {\sc class}~\cite{CLASS1,CLASS2}, or even at second order such as in {\sc  song}~\cite{Pettinari:2013he}. From this, the density and velocity fields of the cosmic fluid are obtained at some high redshift, typically 
in the range $49\lesssim z_{\rm ini}\lesssim99$. In the case of Gaussian initial perturbations, these are fully characterised by the two-point correlation function (or the power spectrum in Fourier space). However, in order to actually use this cosmic fluid data as the ICs for an $N$-body code, it requires a method for mapping this to the particles' phase space. For the following discussion we assume that no vorticity is present at the initial redshift, although this is naturally developed at late times due to the nonlinear evolution.

\subsection{The displacement vector method}\label{sec:IC-displacement-method}

The problem of realising an initial matter power spectrum $P_\Delta(k;z_{\rm ini})$ related to some Gaussian random field $\Delta({\bf x};z_{\rm ini})$ using a particle distribution can be approached in terms of a density-displacement duality. Then the generation of particles' initial positions reduces to the calculation of a displacement vector $\chi^i$ which maps the positions from a regular grid or glass configuration~\cite{Baugh95:glass}, $q^i$, to the perturbed positions, $x^i({\bf q})$, via
\begin{equation}\label{eq:LPT-pos}
    x^i({\bf q})=q^i+\chi^i({\bf q})\,.
\end{equation}
Equation~\eqref{eq:LPT-pos} can be regarded as a coordinate transformation from some virtual coordinate system $q^i$ with constant mass (or charge) density per coordinate volume $\bar{Q}$ to a physical coordinate system $x^i$ where the density field $Q({\bf x})=\bar{Q}\left[1+\Delta({\bf x})\right]$ is inhomogeneous\footnote{Note that here we use $Q$ ($\Delta$) rather than $\rho$ ($\delta$) to denote the density (overdensity) variable since this does not necessarily correspond to the physical $\rho$ appearing in the energy-momentum tensor Eq.~\eqref{eq:energy-momentum-tensor}, as we will see later.}. By virtue of mass conservation, this mapping must satisfy
\begin{equation}\label{eq:LPT-mass-conservation}
    \bar{\rho}{\rm d}^3{\bf q} = \bar{\rho}\left[1+\Delta({\bf x})\right]{\rm d}^3{\bf x}\,.
\end{equation}
Since ${\rm d}^3{\bf x}/{\rm d}^3{\bf q}=\det(J)$, where $J^i_j=\delta^i_j+\partial\chi^i/\partial{q}^j$ is the Jacobian of the transformation (\ref{eq:LPT-pos}), we can expand Eq.~\eqref{eq:LPT-mass-conservation} perturbatively if $|\partial\chi^i/\partial{q}^j|\ll1$. Then, at the linear level we find
\begin{equation}\label{eq:ZA}
    \Delta({\bf x})=-\partial_i\chi^i\,,
\end{equation}
which corresponds to the Zel'dovich approximation \cite{Zeldovich:1969sb} and defines the displacement vector at the initial redshift $z_{\rm ini}$. Consequently, the particles' coordinate velocity can be calculated as 
\begin{equation}\label{eq:LPT-vel}
v^i\equiv\frac{{\rm d}x^i}{{\rm d}t}=\dot{\chi}^i\,,
\end{equation}
where we note that, at first order, the spatial components of the 4-velocity and coordinate velocities coincide, i.e., $u^i=v^i/u^0\approx v^i$. Using Eqs.~(\ref{eq:ZA}) and \eqref{eq:LPT-vel}, it is straightforward to show that $\Delta$ and $v^i$ satisfy the linear continuity equation,
\begin{equation}\label{eq:LPT-continuity}
    \partial_iv^i+\dot{\Delta}=0\,.
\end{equation}
Namely, in this method, the overdensity variable $\Delta$ to which $\chi^i$ is related, and the velocity variable $v^i$, which is calculated from the latter, are generically linked through the `standard' (`or Newtonian') continuity equation,  Eq.~\eqref{eq:LPT-continuity}. For this reason, we can consider $\{\Delta,v^i\}$ as `conjugated' variables. Let us point out that this is the first limitation that this method has when trying to deal with the generation of ICs for general relativistic simulations, since in a completely arbitrary gauge there is no guarantee that the continuity equation Eq.~\eqref{eq:LPT-continuity} would actually hold for the `density' and `velocity' that appear in $T^{\mu\nu}$, and hence the overdensity in such a gauge and the inferred velocity through the displacement vector method are not necessarily conjugated variables. As we will discuss in Section~\ref{sec:nbody-simulations}, this issue is actually not present in Newtonian $N$-body simulations.

Naturally, Eq.~\eqref{eq:LPT-continuity} shows that, in order to calculate the particles' velocity we require not {exactly} information about the initial density fluctuations $\Delta(z_{\rm ini})$ (encoded in $P_\Delta(k;z_{\rm ini})$), but indeed about its time evolution. At linear order, the overdensity can be written in terms of a linear growth factor as $\Delta(k;z)=D_1\Delta(k;z=0)$, with $D_1=1$ at $z=0$, so that the velocity is given by
\begin{equation}\label{eq:nbody-cont-lin}
    \partial_iv^i=-Hf_1\Delta\,,
\end{equation}
where $f_1\equiv{\rm d}\ln{D_1}/{\rm d}\ln{a}$ is the linear growth rate. As a result, we can determine the velocities from Eq.~\eqref{eq:nbody-cont-lin} by using the input density field $\Delta(z_{\rm ini})$ alongside with some numerical values or fit for $f_1$, which depends on the actual model and theory of gravity. In fact, it is well-known that $D_1$ and $f_1$ are in general scale-dependent in modified gravity and dark energy models~\cite{Linder,Narikawa:2009ux}, in which case the simulation particles are displaced along curved trajectories, rather than straight lines, over time \cite{Valkenburg:2015dsa}, even in the linear regime.

\subsection{The gauge correspondence in Newtonian $N$-body simulations}
\label{sec:nbody-simulations}

In this section, we discuss how the ICs generated by the displacement vector method above are consistent with standard Newtonian $N$-body simulations. The reason behind this is that, in a correspondence between Newtonian theory and GR at the linear level, these simulations use mixed gauges~\cite{Chisari:2011iq,Flender:2012-UE,Hwang:2012} -- the density field tracked can be identified as in the synchronous gauge, while the velocity field corresponds to the Newtonian gauge (but note that one loses track of the synchronous gauge when structure formation progresses to the nonlinear regime where particle trajectories cross each other). In order to understand this, let us start with the Newtonian gauge metric
\begin{equation}\label{eq:newt_metric}
    {\rm d}s^2 = -(1+2\psi){\rm d}t^2 + a^2(1-2\phi)\delta_{ij}{\rm d}x^i{\rm d}x^j,
\end{equation}
where $\psi$ and $\phi$ are the gauge-invariant Bardeen potentials~\cite{Bardeen}. In this gauge, the $(00)$, the $(0i)$ and the traceless part of the $(ij)$ components of the Einstein equation in Fourier space are correspondingly given by
\begin{align}
\label{eq:00-newt}k^2\phi + 3\frac{\dot{a}}{a}\left(\dot{\phi}+\frac{\dot{a}}{a}\psi\right) &= -4\pi Ga^2\bar{\rho}\delta^{\rm N}{_{\rm tot}},\\
\label{eq:0i-newt}k^2\left(\dot{\phi}+\frac{\dot{a}}{a}\psi\right) &= 4\pi Ga^2\left(\bar{\rho}+\bar{P}\right)\theta^{\rm N}{_{\rm tot}},\\
\label{eq:ij-newt}k^2\left(\phi-\psi\right) &= 12\pi Ga^2\left(\bar{\rho}+\bar{P}\right)\Theta^{\rm N}{_{\rm tot}},
\end{align}
where $\theta$ is the velocity divergence, $\left(\bar{\rho}+\bar{P}\right)\Theta\equiv-(\hat{k}_i\hat{k}_j-1/3\delta_{ij})\Sigma^i_j$, and we have used a superscript ${\rm N}$ to denote Newtonian-gauge quantities from the energy-momentum tensor {and the subscript `${\rm tot}$' means this is the total contribution from all matter species (the symbols without this subscript denote the corresponding quantities for individual species)}. The Einstein equations \eqref{eq:00-newt}-\eqref{eq:ij-newt} can be combined into
\begin{equation}\label{eq:00-0i-newt}
    k^2\phi = -4\pi Ga^2\bar{\rho}\left[\delta^{\rm N}{_{\rm tot}}+\frac{3H}{k^2}(1+w)\theta^{\rm N}{_{\rm tot}}\right].
\end{equation}
In addition, the continuity equation $\nabla_\mu T^{\mu 0}=0$ and the Euler equation $\nabla_\mu T^{\mu i}=0$ are, respectively, given as \cite{MB95}
\begin{eqnarray}
\dot{\delta}^{\rm N} + \left(1+w\right)\left(\theta^{\rm N}-3\dot{\phi}\right) + 3\frac{\dot{a}}{a}\left(\frac{\delta P}{\delta\rho}-w\right)\delta^{\rm N} &=& 0,\label{eq:cont-newt}\\
\dot{\theta}^{\rm N} + \frac{\dot{a}}{a}\left(1-3w\right)\theta^{\rm N} + \frac{\dot{w}}{1+w}\theta^{\rm N} - \frac{\delta P/\delta\rho}{1+w}k^2\delta^{\rm N} + k^2\Theta^{\rm N} - k^2\psi &=& 0\,.\label{eq:euler-newt}
\end{eqnarray}
In the linear regime, these equations govern the evolution of $\delta^{\rm N}$ and $\theta^{\rm N}$ of each {non-interacting} component of the cosmic fluid independently, although the metric perturbations are sourced by all of them through the Einstein equations \eqref{eq:00-newt}-\eqref{eq:ij-newt}. 
In the case of dark matter (as is the case of $N$-body simulations), we have $\Theta^{\rm N}=w=\dot{w}=\delta P/\delta\rho=0$, {and \eqref{eq:cont-newt} and \eqref{eq:euler-newt} reduce respectively to}
\begin{align}
\dot{\delta}^{\rm N} +\theta^{\rm N}-3\dot{\phi} &= 0,\label{eq:nbody-newt}\\
    \dot{\theta}^{\rm N} + \frac{\dot{a}}{a}\theta^{\rm N} - k^2{\psi} &= 0,\label{eq:nbody-euler}
\end{align}
{where $\dot{\phi}\neq0$ and $\psi\neq\phi$ in general when the Universe is not matter dominated}. Note that Eq.~\eqref{eq:nbody-newt} actually does {not} have the standard form of the continuity equation, Eq.~\eqref{eq:LPT-continuity}, and thus $\{\delta^{\rm N},v^i_{\rm N}\}$ are not `conjugated' variables, i.e., they seem to be incompatible with the displacement vector method. Naturally, in the special case of pure dark matter domination, where $\dot\phi=0$, Eq.~(\ref{eq:nbody-newt}) does take the standard form, and although this renders $\{\delta^{\rm N},v^i_{\rm N}\}$ compatible with the displacement vector method, we note that Eqs.~(\ref{eq:00-0i-newt}, \ref{eq:nbody-newt}, \ref{eq:nbody-euler}) still take different forms from the particle drift and kick equations and the Poisson equation used in traditional Newtonian $N$-body simulations (note that even if $\dot{\phi}=0$ is a good approximation when radiation and the cosmological constant can be both neglected, at late times we generally have $\dot\phi\neq0$). Indeed, it is well-known that $\delta^{\rm N}$ on large scales is different from the density field measured directly from a snapshot of traditional Newtonian $N$-body simulations. One can, nevertheless, still develop a consistent relativistic $N$-body simulation in the Newtonian gauge, by solving these equations, together with any evolution equation for the total anisotropic stress (which is needed to connect $\phi\neq\psi$) within the simulation. This is in principle the same approach as taken by GR simulation codes such as {\it gevolution}~\cite{Adamek:2016zes} or {\sc gramses}.

Next, consider the same set of Einstein equations and matter conservation laws written in terms of synchronous gauge variables. In this gauge, the metric with scalar perturbations is given by
\begin{equation}\label{eq:sync_metric}
    {\rm d}s^2 = -{\rm d}t^2 + a^2(\delta_{ij}+h\delta_{ij}/3+h^{||}_{ij}){\rm d}x^i{\rm d}x^j,
\end{equation}
where $h^{||}_{ij}=(\partial_i\partial_j-\delta_{ij}/3)(h+6\eta)$, {in which $\eta,h$ are the two metric potentials}~\cite{MB95}. In this gauge, the relevant equation is the (0$i$) component of the Einstein equation,
\begin{eqnarray}\label{eq:0i-sync}
k^2\dot{\eta} = 4\pi Ga^2\left(\bar{\rho}+\bar{P}\right){\theta^{\rm S}_{\rm tot}},
\end{eqnarray}
where we have used a superscript ${\rm S}$ to denote synchronous-gauge quantities from the matter sector. The continuity equation in this gauge takes the form
\begin{eqnarray}\label{eq:cont-sync}
\dot{\delta}^{\rm S} + \left(1+w\right)\left(\theta^{\rm S}+\frac{1}{2}\dot{h}\right) + 3\frac{\dot{a}}{a}\left(\frac{\delta P}{\delta\rho}-w\right)\delta^{\rm S} &=& 0\,.
\end{eqnarray}
Using the gauge transformations Eq.~\eqref{eq:delta-gauge-transf} and Eq.~\eqref{eq:uupperi-gauge-transf}, it can be shown that $\delta$ and $\theta$ in synchronous and Newtonian gauges are related by~\cite{MB95}
\begin{eqnarray}
\delta^{\rm N} &=& \delta^{\rm S} - \frac{3H}{2k^2}(1+w)\left(\dot{h}+6\dot{\eta}\right),\label{eq:sync-newt-delta}\\
\label{eq:sync-newt-theta}\theta^{\rm N} &=& \theta^{\rm S} + \frac{1}{2}\left(\dot{h}+6\dot{\eta}\right).\label{eq:sync-newt-delta-2}
\end{eqnarray}
From Eq.~\eqref{eq:sync-newt-delta} and \eqref{eq:sync-newt-delta-2}, it is clear that the combination $\delta+\frac{3H}{k^2}(1+w)\theta$ is gauge invariant, i.e.,
\begin{equation}
\delta^{\rm N}+\frac{3H}{k^2}(1+w)\theta^{\rm N} = \delta^{\rm S}+\frac{3H}{k^2}(1+w)\theta^{\rm S}.
\end{equation}
Therefore Eq.~\eqref{eq:00-0i-newt} can be written as
\begin{align}\label{eq:temp1}
    k^2\phi &= -4\pi Ga^2\bar{\rho}\left[{\delta^{\rm S}_{\rm tot}}+\frac{3H}{k^2}(1+w){\theta^{\rm S}_{\rm tot}}\right].    
\end{align}
In a universe with collisionless dark matter and a cosmological constant (i.e., with negligible contributions from radiation and baryons) we have ${\theta^{\rm S}_{\rm tot}}=0$, so that Eq.~\eqref{eq:temp1} becomes
\begin{equation}\label{eq:nbody-poisson}
    k^2{\phi} = -4\pi Ga^2\bar{\rho}\delta^{\rm S}\,,
\end{equation}
{where $\delta^{\rm S}$ is the density contrast of dark matter only. We recognise Eq.~\eqref{eq:nbody-poisson} as taking the same form as} the standard Poisson equation that is being solved in Newtonian $N$-body simulations to determine the gravitational potential at each time step, and the overdensity variable used as the source {is equal} to that in the synchronous gauge rather than to $\delta^{\rm N}$. Similarly, under the gauge transformation Eq.~\eqref{eq:sync-newt-delta-2}, for dark matter, Eq.~\eqref{eq:cont-sync} becomes
\begin{equation}\label{eq:nbody-cont}
\dot{\delta}^{\rm S} + \theta^{\rm N}=0\,,
\end{equation}
where we have used $\dot{\eta}=0$ which is a consequence of {$\theta^{\rm S}_{\rm tot}=0$ by assuming there is no radiation or baryons} in Eq.~(\ref{eq:0i-sync}). As a result, Eq.~(\ref{eq:nbody-cont}) suggests that{, under the presence of dark matter and $\Lambda$,} synchronous{-gauge} density contrast and the Newtonian{-gauge} velocity\footnote{{Note that this is a slight abuse of terminology as these two gauges have different spatial hypersurfaces and one cannot naturally define a `Newtonian-gauge' velocity in the synchronous gauge~\cite{Flender:2012-UE}. This, however, does not affect the numerical evaluation, and the `$\theta^{\rm N}$' here should be considered as a combination of synchronous-gauge quantities (cf.~Eq.~(\ref{eq:sync-newt-delta-2})) that takes the same value as the Newtonian-gauge velocity $\theta^{\rm N}$ (with the comparison understood to be done for positions in these two gauges that correspond to the same point in the unperturbed background spacetime). In addition, note that the velocity is $v^i_{\rm S}=0$ in the synchronous gauge, and the time variation of the energy density contrast $\delta^{\rm S}$ is only due to deformations in the spatial part of the metric, so that a particle's coordinate in the synchronous gauge, ${x}^i_{\rm S}$, remains constant over time; in contrast, in $N$-body simulations particle coordinates do evolve over time -- this suggests that this `mixed-gauge' view of Newtonian $N$-body simulations is a practical rather than a fundamental one (see also Ref.~\cite{Fidler:2015npa} for a more recent approach to this issue).}} satisfy the {formal} continuity equation, {Eq.~(\ref{eq:LPT-continuity})}, while $\delta^{\rm N}$ and $v^i_{\rm N}$ actually satisfy { Eq.~(\ref{eq:nbody-newt})}. {This fact, together with  {Eq.~\eqref{eq:nbody-euler} (with $\psi=\phi$) and} the Poisson equation Eq.~\eqref{eq:nbody-poisson} suggest that it can then be considered} that it is the pair $\{\delta^{\rm S},v^i_{\rm N}\}$ that is actually solved in Newtonian $N$-body simulations \cite{Flender:2012-UE,Chisari:2011iq}, and $\{\delta^{\rm S},v^i_{\rm N}\}$ are in fact conjugated variables so that the displacement vector method can be consistently applied to generate the ICs for this kind of simulations. Following \eqref{eq:nbody-cont-lin}, for Newtonian $N$-body simulations the velocity can be calculated consistently by solving
\begin{equation}\label{eq:nbody-cont-lin-2}
    \partial_iv^i_{\rm N}=-Hf_1\delta^{\rm S}\,.
\end{equation}
In standard ICs codes such as {\sc 2LPTic}~\cite{Scoccimarro:1997gr,Crocce2006:2LPT} the growth rate is parameterised as $f_1=\Omega_m(a)^{0.6}$~\cite{Lahav:1991}, with $\Omega_m(a)={{\Omega_{m,0}a^{-3}}}/{{(H/H_0)^2}}$, which is valid only for the growth rate of ${\delta}^{\rm S}$ in a $\Lambda$CDM universe and is fully compatible with a standard Newtonian simulation. Moreover, this allows to calculate second order corrections for the displacement vector based on approximations specific for this scenario and which allow to generate the ICs at even lower redshifts, so that the $N$-body system can be evolved for a shorter period of time. 

Before moving to the gauge used in {\sc gramses}, we briefly mention that an alternative approach to interpret Newtonian $N$-body simulations from a relativistic point of view are the recently-proposed $N$-body \cite{Fidler:2015npa} and Newtonian motion~\cite{Fidler:2016tir,Fidler:2017pnb} gauges, in which the coordinate system is defined such that the linearised dynamical equations of GR match the Newtonian counterparts when considering non-relativistic species, and first-order corrections arising from the latter can be consistently included \cite{Fidler:2017ebh,Adamek:2017-early-rad,Fidler-Neutrinos:2019}. Interested readers can find more details from these references. Let us also note that strictly speaking the identification of the synchronous gauge variable $\delta^{\rm S}$ only makes sense in the linear regime before particle stream crossing, and it becomes ill-defined in the nonlinear regime. On the other hand, on the nonlinear, subhorizon, scales, the difference from the Newtonian gauge density perturbation is {suppressed as the gauge difference formally scales with $(H/k)^2$, see Eq.~\eqref{eq:sync-newt-delta}}. This is a subtle point to bear in mind in the approach taken here, while a more sophisticated gauge definition can eliminate it: $N$-body gauge, for example, offers a unified treatment of these two different regimes by stitching together different subhorizon patches of space using a global coordinate system.

\subsection{Initial conditions for {\sc gramses} simulations}\label{sec:IC-gramses-gauge}

Let us now move on to some relevant aspects for the generation of ICs in the CMC-MD gauge. As in the case of the previous subsection, we start by presenting the continuity equation in this gauge, so that we can identify the actual variables being used in {\sc gramses} simulations and assess its compatibility with the displacement vector method. It can be shown that the continuity equation $\nabla_\mu T^{\mu 0}=0$ in the CMC gauge, at linear order, takes the form
\begin{equation}
\dot{\delta}^{\rm C} + \left(1+w\right)\left(\partial_iu^i_{\rm C}-3\dot{\Psi}\right) + 3\frac{\dot{a}}{a}\left(\frac{\delta P}{\delta\rho}-w\right)\delta^{\rm C} = 0,
\end{equation}
where the subscript/superscript ${\rm C}$ is used to denote CMC-MD gauge quantities from the matter sector. If we consider collisionless non-relativistic dark matter, this reduces to
\begin{equation}\label{eq:continuity-eq-cmc-md}
    \dot{\delta}^{\rm C} + \partial_iu^i_{\rm C} - 3\dot{\Psi} = 0\,.
\end{equation}
Clearly, Eq.~\eqref{eq:continuity-eq-cmc-md} resembles the Newtonian-gauge continuity equation Eq.~\eqref{eq:nbody-newt}, and the term $3\dot{\Psi} $ is not present in either the standard form of the continuity equation Eq.~\eqref{eq:LPT-continuity}, or in the mixed-gauge version used in Newtonian simulations, Eq.~\eqref{eq:nbody-cont}. This additional term represents a volume change due to relativistic deformations of space, which can create an under-dense or over-dense region even in the absence of any peculiar motion of matter~\cite{Fidler:2015npa}. In fact, given that an infinitesimal 3-dimensional volume element is distorted, at linear order, by the factor $\sqrt{\gamma}=a^{3}(1-3\Psi)$, from Eq.~\eqref{eq:s_0} we can show that the fluctuations in the conformal density field $s_0$ are given by
\begin{equation}
\delta s_0\equiv\frac{s_0-\bar{s}_0}{\bar{s}_0}=\delta^{\rm C}-3\Psi\,. 
\end{equation}
where as before an overbar denotes background value. This corresponds to a particle number density contrast in the CMC gauge:
\begin{equation}\label{eq:temp3}
    \delta^p  \equiv \delta^{\rm C} - 3\Psi\,,
\end{equation}
with which Eq.~\eqref{eq:continuity-eq-cmc-md} is cast into the standard form of the continuity equation
\begin{equation}\label{eq:temp4}
    \dot{\delta}^{p} + \partial_iu^i_{\rm C} = 0.
\end{equation}
This indicates that $\{\delta^{p},u^i_{\rm C}\}$ are conjugated variables, so that $u^i_{\rm C}$ can be generated by using Eq.~\eqref{eq:temp4} with $\delta^{p}$ as an input. In this situation, $\delta^p$ can be regarded as fluctuations in the `bare' density field~\cite{Valkenburg:2015dsa} as the spacetime curvature is not included, and at the linear level this corresponds to the perturbations in the conformally-scaled density $s_0$ which is used in {\sc gramses}. For an $N$-body simulation, using $s_0$ rather than $\rho$ is more convenient in practice since we are interested in following `particles' rather than the {total `energy density field'} itself, and the (non-conformal) energy density field at a given instant can be separately calculated by inverting Eq.~\eqref{eq:s_0} {once the spatial metric ${\gamma}_{ij}$ is solved}.

Notice, however, that the identification of $\delta^p=\delta s_0$ is only made at linear order in perturbation theory, which is sufficient for the purpose of setting up ICs. At the nonlinear level, the density field in Eq.~\eqref{eq:s_0} contains additional contributions because $\rho=(\alpha u^0)^2\rho_0$, where $\rho_0$ is the actual rest-mass of the system. The extra factor $\alpha u^0$ actually corresponds to a Lorentz factor which contributes a quadratic term in the velocity that does not affect the generation of ICs at the linear level. In {\sc gramses}, $s_0$ is constructed in a completely nonlinear way using the particles' positions and velocities in a Cloud-in-Cell (CIC) scheme, as well as the updated values of the metric components.

Interestingly, it can be shown that using the gauge transformations Eqs.~\eqref{eq:cmc_psi}, \eqref{eq:h_ij-gauge-transform} and \eqref{eq:delta-gauge-transf}, as well as the linearised MD condition Eq.~\eqref{eq:MD-linear-gauge-condition}, we can rewrite the right-hand side of Eq.~\eqref{eq:temp3} in terms of synchronous gauge variables as (bear in mind that $\delta^p$ is the particle number density contrast in the CMC gauge)
\begin{equation}\label{eq:delta-p-sync}
\delta^p =\delta^{\rm S} - 3\eta\,.
\end{equation}
We note that, while $\dot{\eta}=0$ in the dark-matter and $\Lambda$ dominated eras, $\eta$ is in general not zero, which means that $\delta^p$ and $\delta^{\rm S}$ differ by a scale-dependent function whose shape has been fixed by its evolution at higher redshifts, when the contribution of radiation cannot be neglected.
Since it can be shown that $u^i_{\rm C}=u^i_{\rm N}$ (with contravariant index), as discussed in Section~\ref{sec:gauge-matter-sector}, the fact that Eq.~\eqref{eq:nbody-cont} and Eq.~\eqref{eq:temp4} have the same form is consistent with $\delta^p$ differing from $\delta^{\rm S}$ only by a time-independent quantity. 

Notice, however, that Eq.~\eqref{eq:temp4} allows to calculate the 4-velocity $u^i$ (with upper index) rather than $u_i$, and it is the latter that is the actual variable which appears in the $3+1$ form of the geodesic equation written as a first-order system which is implemented in {\sc gramses}~\cite{baumgarte2010:NR-book}. For non-relativistic particles, this is given at linear order by
\begin{align}
\frac{{\rm d}x^i}{{\rm d}t} &= a^{-2}\delta^{ij}u_j - a^{-2}\delta^{ij}\beta_i,\label{eq:geodesic-1}\\
\frac{{\rm d}u_i}{{\rm d}t} &= - \partial_i\Phi\,,\label{eq:geodesic-2}
\end{align}
where the shift vector $\beta^i$ appears explicitly due to to the relation
\begin{equation}\label{eq:uupperi-uloweri}
u_i = a^{2}\delta_{ij}\left(u^j +\beta^j\right)\,.
\end{equation}
The correction due to the shift vector in Eq.~\eqref{eq:uupperi-uloweri}, which in the MD gauge is given by Eq.~\eqref{eq:MDC-CF}, is taken into account in {\sc gramses} itself when starting the simulation by solving the linearised version of the momentum constraint for $\beta^i$ at $z_{\rm ini}$ {(just for once)}, i.e.,
\begin{equation}\label{eq:beta-IC}
({\bar\Delta}_L{\beta}_{\rm ini})^i-6{\Omega_m}\beta^i_{\rm ini}=6\frac{\Omega_m}{a}{u}^i_{\rm ini}\,,
\end{equation}
such that the initial lower-index 4-velocity $u^{\rm ini}_i$ can be constructed using Eq~\eqref{eq:uupperi-uloweri}, only after which does the actual simulation start. {Note that this means that the particle velocity fed into {\sc gramses} is $u^i$ rather than $u_i$.}

Even though we have shown that $\{\delta^{p},u^i_{\rm C}\}$ are conjugated variables, and that we can calculate $u_i^{\rm C}$ using Eq.~\eqref{eq:uupperi-uloweri} and Eq.~\eqref{eq:beta-IC}, there is still one remaining caveat which is the fact that the growth rate of the number density perturbation $\delta^p$ is not known, so that if we insist on using the analogous of \eqref{eq:nbody-cont-lin} to calculate $u^i_{\rm C}$, then $f_1$ needs to be found numerically by solving the evolution equation for $\delta^p$ that can be derived from the corresponding Euler equation and Eq.~\eqref{eq:temp4}. This is doable, but it is not the approach that we shall follow in this paper. 

{Another alternative approach to generate simulation ICs for the CMC gauge is by exploiting the relation $u^i_{\rm C}=u^i_{\rm N}$, and splitting the generation of ICs into two steps: in the first we generate particle displacements using the power spectrum of $\delta^{p}$ as input, while in the second we generate the velocity {$u^i_{\rm N}$} by applying the standard method based on Eq.~(\ref{eq:nbody-cont-lin}), using the power spectrum of $\delta^{\rm S}$ (which has scale-independent linear growth) rather than $\delta^{p}$.
This actually is a neat and simpler way than the more general method presented in Section~\ref{sec:FD-method}, and there is a potential of applying similar tricks to generate ICs in other gauges (though we shall not discuss this further in this paper). We have checked that the relative difference between the velocity divergence power spectra of the ICs generated by using these two methods is well below $0.1\%$.}

\subsection{A finite difference method for the calculation of initial velocities}
\label{sec:FD-method}

To avoid using an explicit parameterisation of $f_1$ for the calculation of velocities, we introduce a finite difference approach. Here, instead of using a single power spectrum $P_{\Delta}(k;z_{\rm ini})$ at $z_{\rm ini}$ as in the standard displacement vector method based on the Zel'dovich approximation, we use two additional power spectra $P_{\Delta}(k;z_{\pm})$ from the neighbouring redshifts $z_{\pm}=z_{\rm ini}\pm\Delta z$, with $\Delta z\ll z_{\rm ini}$ which will provide the information needed about the growth rate of density perturbations around $z_{\rm ini}$. Then, rather than using Eq.~\eqref{eq:LPT-continuity} or Eq.~\eqref{eq:nbody-cont-lin} to calculate the initial velocity, we take the finite difference $\Delta x^i\equiv x^i({z_-})-x^i({z_+})$ directly from the definition of coordinate velocity in Eq.~\eqref{eq:LPT-vel}, i.e.
\begin{equation}\label{eq:LPT-mod-vel}
v^i\approx\frac{\Delta x^i}{\Delta t}=\frac{H}{a}\frac{\chi^i({z_-})-\chi^i({z_+})}{2\Delta z}\,,
\end{equation}
where $\chi^i({z_\pm})$ are the displacement vectors calculated from $P_{\Delta}(k;z_{\pm})$ via the Zel'dovich approximation Eq.~\eqref{eq:ZA} at the neighbouring redshifts $z_{\pm}$. The advantage of using Eq.~\eqref{eq:LPT-mod-vel} over Eq.~\eqref{eq:nbody-cont-lin} is that this approach is independent of the underlying theory of gravity or gauge used since this information is entirely drawn from the input power spectra (obtained from a suitable linear Boltzmann code for the model), and can be applied as long as the density perturbations and velocity needed are conjugated variables. {Importantly, to create the random realisations of $\chi(z_{\rm ini})$ and $\chi(z_{\pm})$ from $P_{\Delta}(k;z_{\rm ini})$ and $P_{\Delta}(k;z_{\pm})$, the same random number sequence, and hence random number seeds, should be used to ensure that we have generated three consecutive snapshots for the `same' particles.}

The fact that the velocity calculation in Eq.~\eqref{eq:LPT-mod-vel} relies entirely on the input power spectra to generate the ICs can potentially become problematic due to the generic presence of radiation at $z_{\rm ini}$ in the linear code, which contributes to the growth rate of matter and drives $\dot\eta$ away from zero. In order to suppress this effect, the two neighbouring power spectra $P_\Delta(k;z_{\pm})$ can be calculated by evolving $P_\Delta(k;z_{\rm ini})$ under the linear theory assuming {matter domination}. We will show how this can be done in particular for the case of {\sc gramses} ICs in Section~\ref{sec:results}.

\begin{figure}
\centering
\includegraphics[width=1\linewidth]{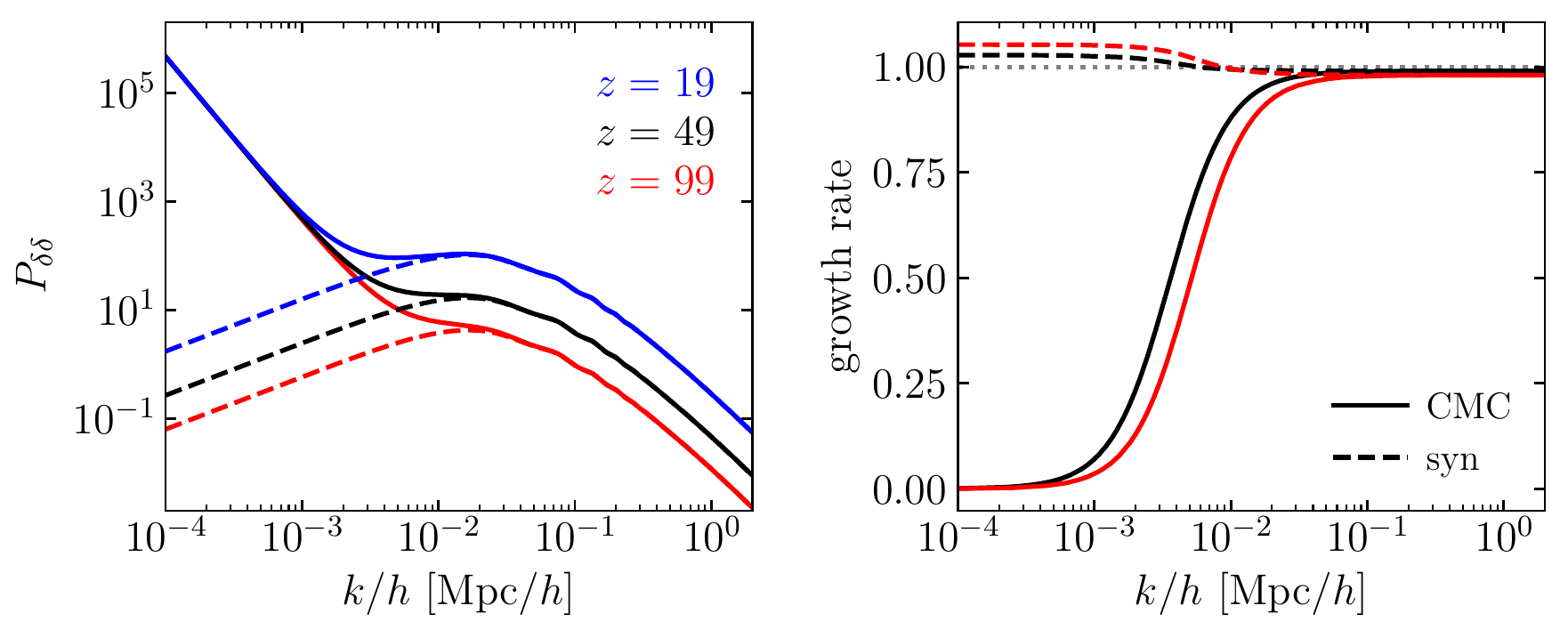}
\caption{{\it Left panel}: a comparison of matter power spectra in the CMC-MD gauge (solid line) and synchronous gauge (dashed line) for three different redshifts -- $z=99$ (red), $49$ (black) and $19$ (blue) -- as obtained from a modified version of {\sc camb}. {\it Right panel}: the growth rates in these two gauges calculated by finite difference of two neighbouring power spectra around the three aforementioned redshifts with $\Delta z=0.5$, see Eq.~\eqref{eq:growth-rate-FD}. As a reference, the gray dotted line shows the linear-theory prediction $f_1=1$ for synchronous gauge in a matter dominated universe.}
\label{fig:2LPTic-ICs-comp}
\end{figure}

Figure~\ref{fig:2LPTic-ICs-comp} illustrates the gauge effects (for synchronous and CMC gauges) on the matter power spectrum (left panel), as well as the scale dependence of $f_1$ due to radiation in these two gauges (right panel). Here, the growth rates of $\delta^p$ and $\delta^{\rm S}$ are obtained by taking finite difference between snapshots in each gauge. Since in Fourier space $P_{\delta\delta}(k;z)=|\delta(k;z)|^2$, this is given by
\begin{equation}\label{eq:growth-rate-FD}
    f^{\rm FD}_1(z)=\frac{1}{2a\Delta z}\frac{\sqrt{P_{\delta\delta}(z_{-})}-\sqrt{P_{\delta\delta}(z_{+})}}{\sqrt{P_{\delta\delta}(z)}}\,,
\end{equation}
where the {\rm FD} superscript means that the left-hand side quantity has been obtained using a finite difference. In this we use the power spectra obtained from a modified version of {\sc camb}, {which works for different gauges}. From the right panel of Fig.~\ref{fig:2LPTic-ICs-comp} we can see that, even for synchronous gauge, the presence of radiation has a noticeable effect at large scales and boosts the growth rate by $\sim2\%$ with respect to the linear-theory prediction for a matter dominated universe ($f_1=1$), and there is an approximately $1\%$ suppression on sub-horizon scales, where gauge effects are not present. For the CMC-MD case, the gauge effects on the overdensities and the scale dependence of the growth rate are also evident from Fig.~\ref{fig:2LPTic-ICs-comp}, as we can see that the power spectra and growth rates agree in both gauges at scales inside the horizon, but there is a dramatic suppression of the growth rates toward large scales in the CMC-MD gauge. This effect arises due to the $\eta$ in Eq.~\eqref{eq:delta-p-sync}, which dominates the shape of the power spectrum below $k\lesssim10^{-3}h{\rm Mpc}^{-1}$, as can be seen from the left panel of Fig.~\ref{fig:2LPTic-ICs-comp}. Since $\eta$ does not evolve considerably on the redshift range shown, at very large scales the matter power spectrum of $\delta^p$ remains roughly constant in time, resulting in the strongly suppressed growth rate depicted in the right panel of Fig.~\ref{fig:2LPTic-ICs-comp}.

For Newtonian simulations, a common approach to take into account the presence of radiation, at least at the background level, is through the so-called `back-scaling' method. In this approach, the input power spectrum used to generate the ICs is given by the linear code at $z=0$ rather than at the actual starting redshift of the simulation. Then, this power spectrum is evolved backwards, up to $z_{\rm ini}$, using a growth factor derived from Newtonian theory (where no radiation is present), resulting in a matter power spectrum $\tilde{P}_{\Delta}(k;z_{\rm ini})$ that does not agree with the Boltzmann code at $z_{\rm ini}$ but that allows to generate `artificial' ICs that guarantees that the simulation will reproduce the correct matter overdensities on linear scales at $z=0$ due to this particular calibration. An alternative, relativistic back-scaling approach has been discussed in \cite{Fidler:2017ebh,Adamek:2017-early-rad}, in which the resulting output of the simulation is interpreted in terms of a different gauge with the aid of a modified version of {\sc class}. 

The finite-difference method described above can in principle be used in combination with the `back-scaling' method, with the latter providing the linear power spectra at not just $z_{\rm ini}$ but also $z_\pm$.
However, our ultimate goal is to start from a more accurate IC, while also include radiation effect in the simulation itself, e.g., by treating radiation as linear perturbations and interfacing with a Boltzmann code such as {\sc camb} or {\sc class} during the $N$-body simulation to calculate relevant quantities. The latter is what we plan to do in the future simulations, which is another reason why in this paper we do not follow the back-scaling approach.

\section{Results}
\label{sec:results}

In this section, we present results on the generation of ICs for {\sc gramses} using the finite difference method described in Section~\ref{sec:FD-method}. We have implemented this method in {\sc 2LPTic} code, so that the initial particles position are calculated with the Zel'dovich approximation (\ref{eq:ZA}) at $z_{\rm ini}$, as in the default {\sc 2LPTic} code\footnote{Contrary to the default {\sc 2LPTic} code, however, we do not use the `back-scaling' of a $z=0$ input power spectrum with a growth factor parameterisation but directly use the one generated by the Boltzmann code at $z_{\rm ini}$.}, but their velocities are calculated using the finite difference expression (\ref{eq:LPT-mod-vel}). For all realisations of the density field we use the same random seed in order to suppress realisation scatter in our results. As previously mentioned, the default {\sc 2LPTic} code uses second-order corrections for the calculation of the displacement vector, while our method implements only linear perturbations as this is enough for the purpose of fixing the gauge issues and generating ICs for {\sc gramses} consistently, so these are turned off for comparison. Initial conditions generated by this method have been used to run the first {\sc gramses} cosmological simulations discussed in the main paper, Ref.~\cite{Barrera-Hinojosa:2019mzo}. The input matter power spectra for the ICs generation are obtained from a modified version of {\sc camb} implementing Eq.~\eqref{eq:delta-p-sync} to relate $\delta^p$ to the synchronous-gauge overdensity $\delta^{\rm S}$ which is the default variable used in such code.

Let us add some details on how we address the problem of radiation in the generation of ICs for {\sc gramses}. For this particular case, we can use Eq.~\eqref{eq:delta-p-sync} to enforce the $\dot{\eta}=0$ condition via
\begin{equation}\label{eq:delta-p-lin}
    \delta^{p}(k;z_{\pm})=\frac{D_1(z_{\pm})}{D_1(z_{\rm ini})}\delta^{\rm S}(k;z_{\rm ini})-3\eta(k;z_{\rm ini})\,,
\end{equation}
where on the right-hand side $\eta$ is {a scale-dependent function constant in time, evaluated at $z_{\rm ini}$}. Then, it is sufficient to calculate the linear growth factor in the synchronous gauge, $D_1(z)$, with which $\delta^{\rm S}$ can be evolved from $z_{\rm ini}$ to $z_{\pm}$ in the absence of radiation while keeping $\eta$ fixed by outputting $\eta(z_{\rm ini})$ from {\sc camb}. As a result, $P_{\delta\delta}(k;z_{\pm})=|\delta^p(k;z_{\pm})|^2$ can be constructed from \eqref{eq:delta-p-lin} in such a way that it is completely free from radiation effects and allows for generating the velocities for the {\sc gramses} $N$-body simulation using Eq.~(\ref{eq:LPT-mod-vel}). 

There is an additional subtlety to take into account in order to compare the ICs from this method against the ones generated from default {\sc 2LPTic}, which is that the normalisation\footnote{By `normalisation' here we mean the root-mean-squared fluctuation of matter smoothed on $8h^{-1}$Mpc scales, $\sigma_8$. In 2{\sc lpt}ic code, the value of $\sigma_8$ is required as an input to get the correct amplitude of the initial matter density field.} of the matter power spectra $P_\Delta(k;z_{\pm})$ is required as an input (while the default {\sc 2LPTic} code only requires this at $z_{\rm ini}$). These have associated values of $\sigma_8$ at the neighbouring redshifts $z_{\pm}$ and directly affect the calculation of the displacement vectors and hence the velocity via Eq.~\eqref{eq:LPT-mod-vel}. Therefore, even the Newtonian velocity calculated by {the finite difference} method will not necessarily coincide with what is calculated by using the growth rate $f_1$ parameterisation implicit in the default {\sc 2LPTic} code. In order to make these comparable, we can get a linear-theory prediction for the $\sigma_8$ values at the neighbouring redshifts by applying the linear growth rate as 
\begin{equation}
\sigma^{\rm S}_8(z_{\pm})=\frac{D_1(z_{\pm})}{D_1(z_{\rm ini})}\sigma^{\rm S}_8(z_{\rm ini})\,,
\end{equation}
so we can use these rather than the {\sc camb} output values at $z_\pm$.

Similarly, in the CMC gauge, we can use Eq.~\eqref{eq:delta-p-lin} to estimate $\sigma^p_8$ from linear theory: its values at the neighbouring redshifts with fixed $\eta$ can be calculated as
\begin{equation}\label{eq:sigma8-lin}
    \sigma^p_8(z_{\pm})=\sqrt{\left[\frac{D_1(z_{\pm})}{D_1(z_{\rm ini})}\sigma^{\rm S}_8(z_{\rm ini})\right]^2 +9[\sigma^{\eta}_8(z_{\rm ini})]^2 -6\frac{D_1(z_{\pm})}{D_1(z_{\rm ini})}\left[\sigma^{\rm{S}\eta}_8(z_{\rm ini})\right]^2}\,,
\end{equation}
where $\sigma^{\eta}_8$ represents the normalisation of the power spectrum $P_{\eta\eta}$, and $\sigma^{\rm{S}\eta}_8$ that of the cross spectra between $\delta_{\rm{S}}$ and $\eta$. We calculate this cross-term by evaluating Eq.~\eqref{eq:sigma8-lin} at the initial redshift $z_{\rm ini}$, i.e.,
\begin{equation}\label{eq:sigma-cross-lin}
    [\sigma^{\rm{S}\eta}_8(z_{\rm ini})]^2=\frac{[\sigma^{\rm S}_8(z_{\rm ini})]^2-[\sigma^p_8(z_{\rm ini})]^2+9[\sigma^{\eta}_8(z_{\rm ini})]^2}{6}\,.
\end{equation}
This way, we can ensure that the power spectra normalisation at the neighbouring redshifts is consistent with those constructed from Eq.~\eqref{eq:delta-p-lin}. 

\begin{figure}
    \centering
    \includegraphics[width=\linewidth]{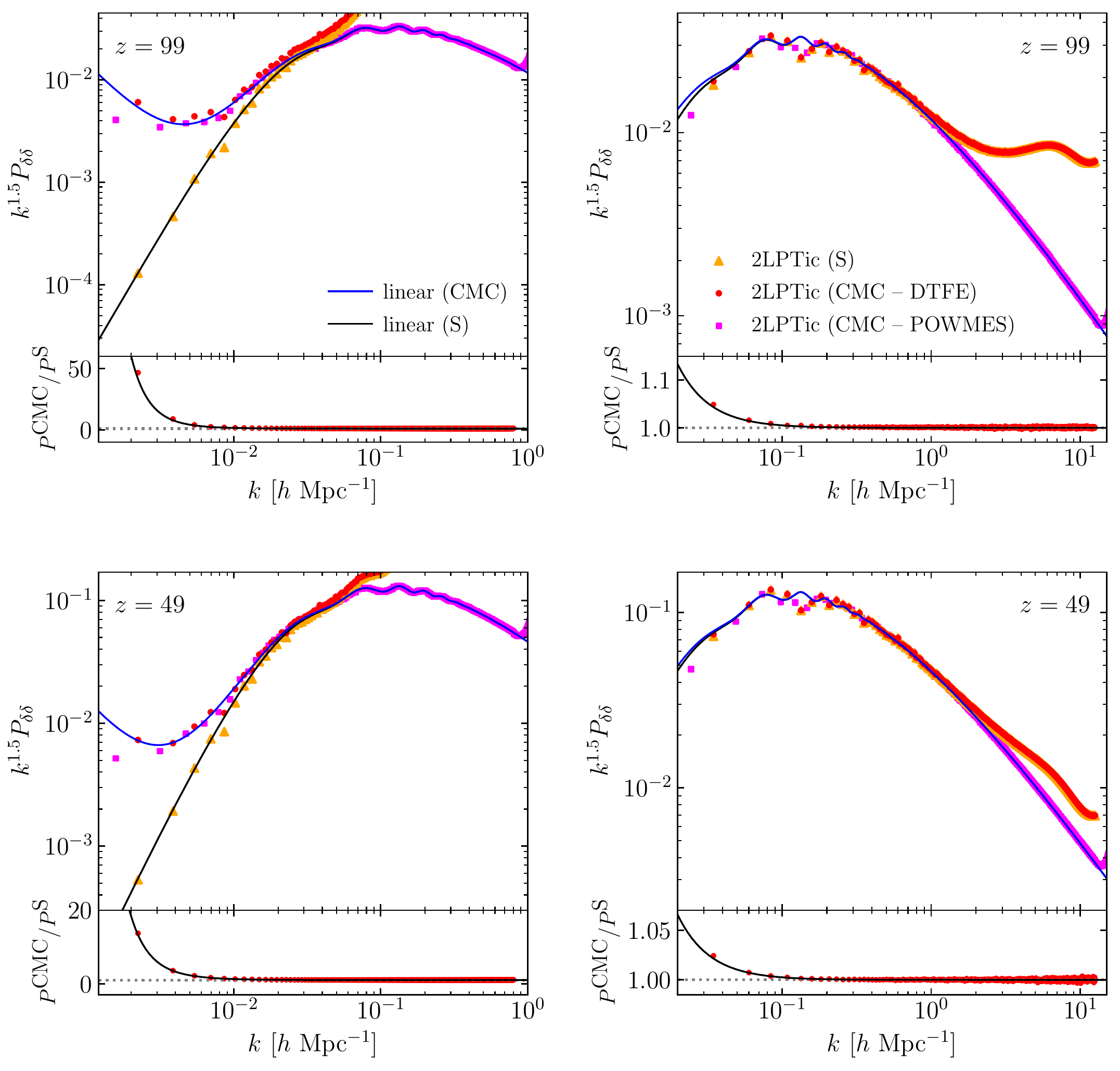}
    \caption{Matter power spectra of ICs generated by both standard and modified {\sc 2LPTic} codes. {\it Left panels}: results for the low-resolution setup ($L=4h^{-1}$Gpc and $N_{p}=1024^3$) for $z_{\rm ini}=99$ (top) and $z_{\rm ini}=49$ (bottom). {\it Right panels}: results for the high-resolution setup ($L=256h^{-1}$Mpc and $N_{p}=1024^3$) for $z_{\rm ini}=99$ (top) and $z_{\rm ini}=49$ (bottom). In all panels, the linear-theory predictions for synchronous and CMC gauges are represented by the solid black and blue lines, respectively. The red circles in the bottom sub-plots represent the ratio between $P_{\delta\delta}$ in the CMC-MD and synchronous gauges as measured by {\sc dtfe} code, while the solid black line represents the ratio between the linear-theory curves.}
    \label{fig:2LPTic-IC-mat}
\end{figure}

In order to illustrate how well this method works, we generate the ICs for two different setups; a low-resolution one with a comoving box size of $L=4h^{-1}$Gpc and $N_{p}=1024^3$ dark matter particles, and a high-resolution one with $L=256h^{-1}$Mpc and the same number of particles. We also use two different initial redshifts, and {the power spectra measured from these ICs} are compared against linear-theory predictions obtained from {the modified} {\sc camb} version.

Figure~\ref{fig:2LPTic-IC-mat} shows the matter power spectrum of the ICs generated by both the standard and modified version of {\sc 2LPTic}. In the case of the CMC-MD gauge, this is measured from the ICs data in two different ways; one is using the {\sc dtfe} code~\cite{Cautun:2011-DTFE} along with {\sc nbodykit}~\cite{nbodykit:2018} (red circles), and the second one is applying {\sc powmes}~\cite{Colombi:2008dw} (magenta squares). The reason for this is that the former captures more accurately the turnover on large scales due to gauge effects but might lose accuracy towards shorter wavelength modes, while {\sc powmes} has a better performance when approaching the Nyquist frequency but eventually fails at capturing the largest-scale components appearing in this particular gauge. For the synchronous gauge case (orange triangles) only {\sc DTFE} and {\sc nbodykit} are applied. Figure~\ref{fig:2LPTic-IC-mat} shows good agreement between the linear-theory predictions and the {\sc 2LPTic} results for both the high and low resolution setups and both initial redshifts, although {the latter} decreases towards larger scales where gauge effects dominate. On these scales we can also see that the ICs data seem to mismatch the linear theory prediction curve, but this is normally the case due to realisation scatter and cosmic variance. However, since the ICs have been generated using the same initial random seeds in all cases, this effect is removed in the ratio between the power spectra in the two gauges, as can be seen from the lower subpanels of each panel.

\begin{figure}
    \centering
    \includegraphics[width=\linewidth]{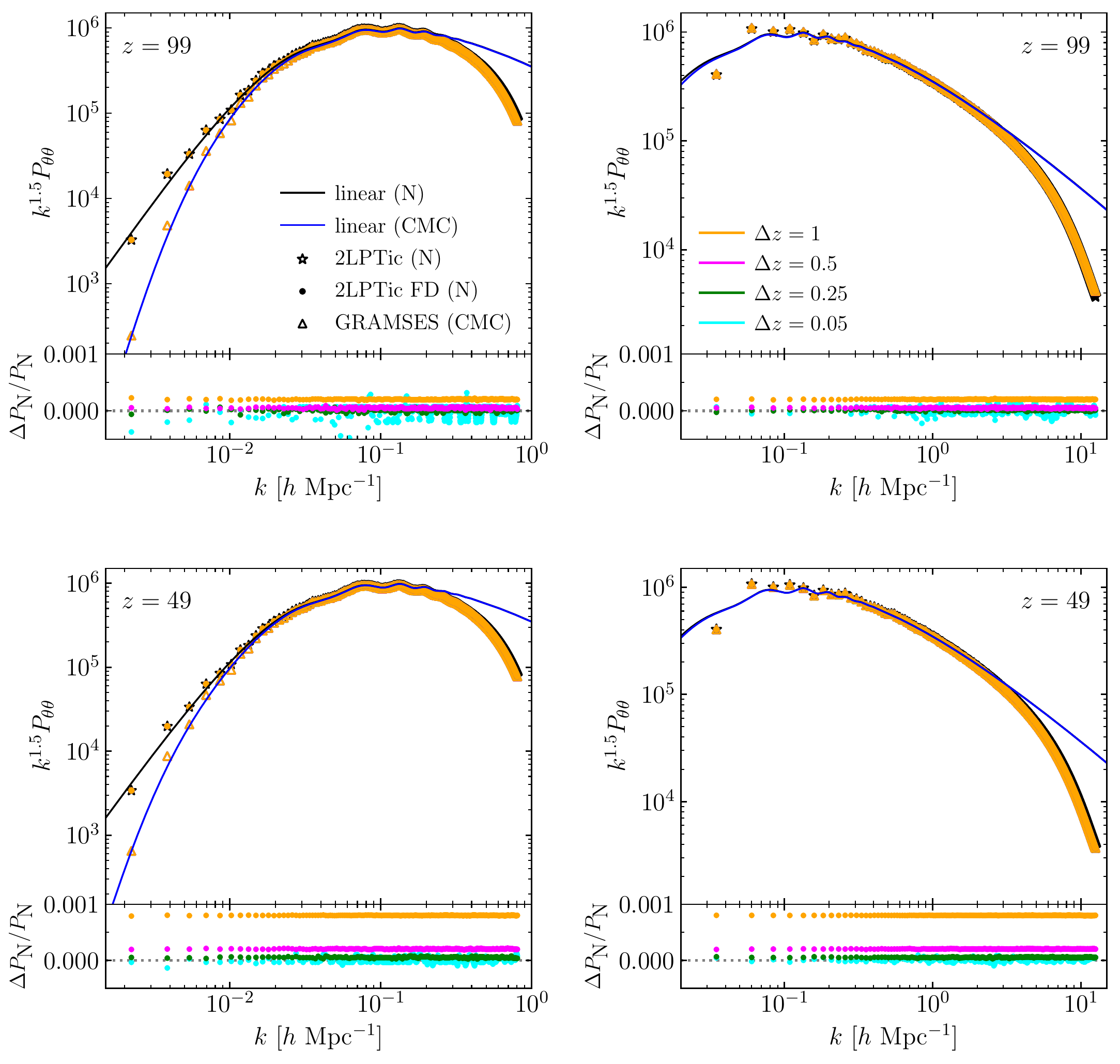}
    \caption{Velocity divergence ($\theta\equiv\nabla\cdot u$) power spectra for the Newtonian gauge measured from the ICs generated by the standard {\sc 2LPTic} code (stars) and its modified version (circles), as well as the CMC-MD result obtained from the {\sc gramses} correction (triangles) discussed in the main text. {\it Left panels}: results for the low-resolution setup ($L=4h^{-1}$Gpc and $N_{p}=1024^3$) for $z_{\rm ini}=99$ (top) and $z_{\rm ini}=49$ (bottom). {\it Right panels}: results for the high-resolution setup ($L=256h^{-1}$Mpc and $N_{p}=1024^3$) for $z_{\rm ini}=99$ (top) and $z_{\rm ini}=49$ (bottom). In all panels, the linear-theory predictions for Newtonian and CMC-MD gauges are represented by the solid black and blue lines, respectively. The bottom sub-panels of each plot show the relative difference between the Newtonian gauge results for the different values of $\Delta z$.}
    \label{fig:2LPTic-IC-vel}
\end{figure}

Figure \ref{fig:2LPTic-IC-vel} is similar to Fig.~\ref{fig:2LPTic-IC-mat}, but shows the results for the velocity divergence $\theta\equiv\nabla\cdot u$ power spectra of the ICs generated using the original {\sc 2LPTic} code (stars) as well as the modified version implementing the finite difference calculation for the velocity field Eq.~\eqref{eq:LPT-mod-vel} for both the Newtonian gauge (circles) and CMC-MD gauge (triangles). In all cases, $\theta$ is calculated using the {\sc dtfe} code. As Eq.~\eqref{eq:LPT-mod-vel} depends on $\Delta z$, we have tried four different values of $\Delta z$ -- $1, 0.5, 0.25, 0.05$ -- to assess the correct magnitude to be used and how this affects the calculation of the velocity field. In the lower subpanels of each panel, we show the relative difference between the Newtonian velocity divergences $\partial_iu^i_{\rm N}$ obtained from the modified and {default} {\sc 2LPTic} code, where sub-percent differences are found for all probed scales and all $\Delta z$ values used, and the amplitude seems robust against spatial resolution. This result suggests that, at least for the case of Newtonian gauge, the finite difference method can be used to generate the ICs for cosmological simulations regardless of specifications and matching the default {\sc 2LPTic} code accuracy (at least up to first order). Nonetheless, we notice that using a value of $\Delta z$ that is too small might introduce some scatter, while increasing $\Delta z$ will monotonically increase the amplitude of the relative difference so that this might eventually become unacceptable (e.g., larger than $\mathcal{O}(1\%)$). From Fig.~\ref{fig:2LPTic-IC-vel} we note that $\Delta z/z_{\rm ini}\sim 1\%$ is enough to suppress noise while keeping the relative difference with respect to the default code under $0.03\%$ in all cases.

As we have remarked before, the ICs method generates $u^i$ while $u_i$ (lower index) is the actual variable used to solve the geodesic equation in the standard $3+1$ (ADM) form {implemented} in the simulations. Thus, in Fig.~\ref{fig:2LPTic-IC-vel} the results for the $\theta_{\rm C}=\partial^i u^{\rm C}_i$ spectra (triangles) have been obtained from the initial {\sc gramses} snapshot, which is outputted after the code solves Eq.~\eqref{eq:beta-IC} to get the shift vector $\beta^i$ and calculates $u^{\rm C}_i$ from Eq.~\eqref{eq:uupperi-uloweri} to start the actual simulation with. {This is how {\sc gramses} operates, so that only $u^{i}$ is needed as an input in the ICs.} In this case, we also find good agreement with the linear-theory expectations for both simulation setups (left and right panels) and redshifts (top and bottom panels), and the deviation from linear theory at small scales (which is also present in the Newtonian gauge cases) is due to spatial resolution effects. 

\begin{figure}
    \begin{subfigure}[b]{1\textwidth}
    \centering
    \includegraphics[width=\linewidth]{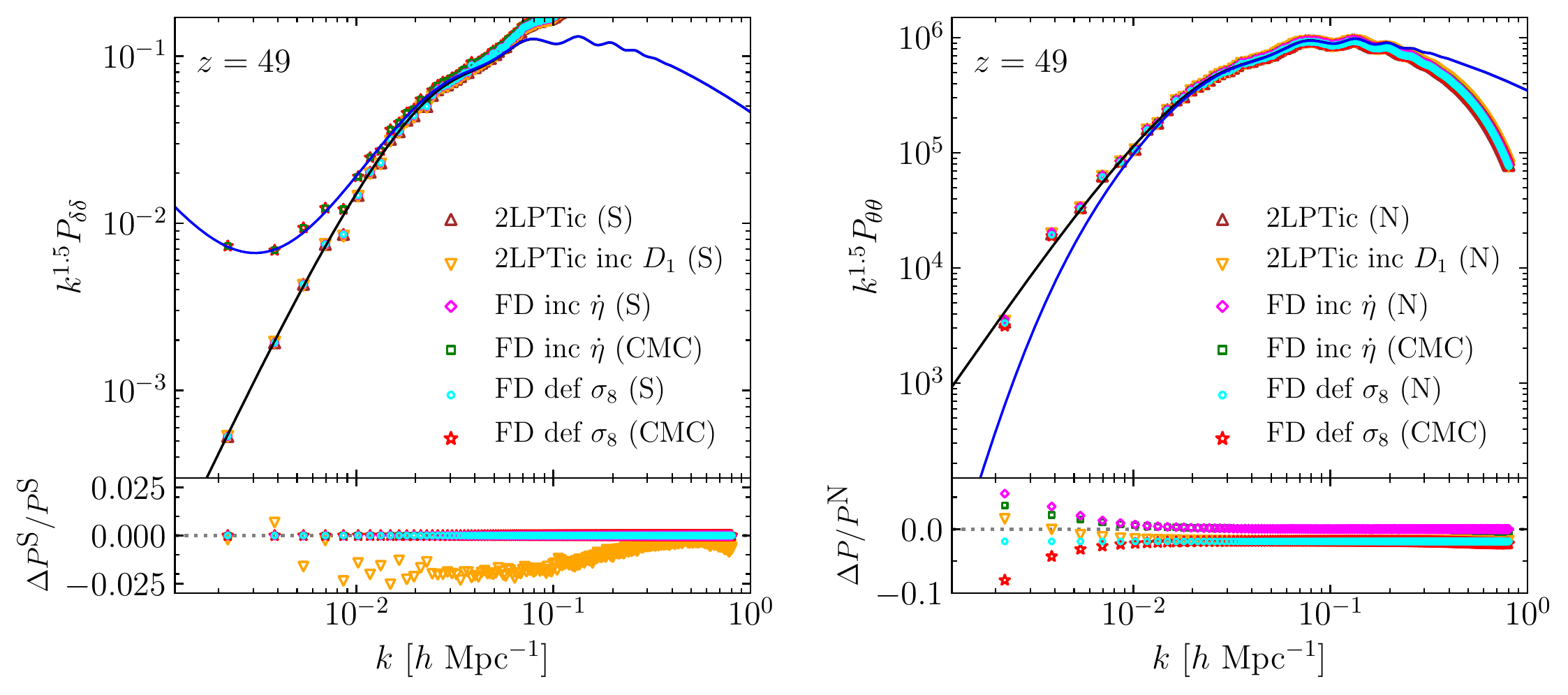}
    \end{subfigure}
    \caption{The impacts of radiation and $\sigma_8$ normalisation on the generation of ICs. The results from the default 2{\sc lpt}ic code, where the linear growth factor $D_1$ is used to `back-scale' the $z=0$ matter power spectrum to $z_{\rm ini}$, is also included (orange triangles). The default case (brown triangles), which is used as the denominator to calculate the relative differences in the lower subpanels, also uses the standard 2{\sc lpt}ic code but feeds it with the matter power spectrum at $z_{\rm ini}$ and sets $D_1=1$. {\it Left panel}: matter power spectra for the overdensity in the synchronous (S) gauge, $\delta^{\rm S}$, as well as for the particle number overdensity $\delta^p$ in the CMC gauge.
    {\it Right panel}: velocity divergence ($\theta=\partial_iu^{i}$) power spectra, where $u^i$ represents the conjugated variable to either $\delta^{\rm S}$ (N) or $\delta^p$ (CMC); note that in the case of the CMC gauge the velocity divergence is not $\partial^iu^{\rm C}_i$. The solid curves denote the linear-theory predictions using the same colours as the previous figures. See the main text for more details of the symbols.}
    \label{fig:ICs-no-corrections}
\end{figure}

Finally, to briefly illustrate the impact of radiation effects and the $\sigma_8$ normalisation on the generation of ICs, in Fig.~\ref{fig:ICs-no-corrections} we show the results when linear-theory corrections are not applied, i.e., when skipping Eqs.~\eqref{eq:delta-p-lin}-\eqref{eq:sigma-cross-lin}, as well as the case where the `back-scaling' method of the standard 2{\sc lpt}ic is used (orange triangles). In the latter case, a $z=0$ power spectrum from {\sc camb} is provided as an input and the code uses the theoretical value of $D_1$ to scale it back to $z_{\rm ini}$. Both panels in Fig.~\ref{fig:ICs-no-corrections} show the results from the standard 2{\sc lpt}ic code with no `back-scaling' (i.e., feeding the 2{\sc lpt}ic code with the linear power spectrum at $z_{\rm ini}$ and setting $D_1=1$; brown triangles) and from the finite difference method for the cases where either: 1) $\eta$ is not kept constant for the default gauge used in 2{\sc lpt}ic (magenta diamonds) and the CMC gauge (green squares), and 2) when $\sigma_8$ is not corrected for the default gauge (cyan circles) and the CMC gauge (red stars). In this case we use the setup with $L=4h^{-1}$Gpc and $N_{p}=1024^3$, and $\Delta z=0.5$ for the finite difference calculations.

In the left panel of Fig.~\ref{fig:ICs-no-corrections}, we show $P_{\delta\delta}$ for all cases in the top panel, and the relative differences with respect to the default case (represented by the brown triangles) are shown in the bottom panel. From the latter, we can see that the back-scaled case (orange triangles) shows a suppression of up to $\sim2.5\%$, which is because radiation and baryons are present in the forward linear-theory calculation (by {\sc camb}) all the way down to $z=0$, while $D_1$ is calculated by assuming a universe with only dark matter and cosmological constant (so that there is an inconsistency). As expected, the synchronous-gauge results for the cases where $\dot{\eta}\neq0$ (magenta diamonds) and where $\sigma_8$ is not corrected for (cyan circles) agree perfectly with the default case, since these corrections apply only to the neighbouring snapshots used for the velocity calculation (via finite difference), while the particle displacements in the ICs are obtained by using solely the matter power spectrum at $z_{\rm ini}$, which are not affected by these corrections.

In the right panel of Fig.~\ref{fig:ICs-no-corrections}, we present the power spectra for the velocity divergences associated to the overdensity variables for all cases shown in the left panel in the top, and the relative differences of $P_{\theta\theta}$ with respect to the default case (Newtonian-gauge velocity obtained by using the 2{\sc lpt}ic code without doing the $D_1$ `back-scaling'; brown triangles) are plotted in the bottom. By comparing the various Newtonian-gauge results it can be seen that, when the $\sigma_8$ correction is not used in the finite difference calculation (cyan circles) a constant $\sim2\%$ suppression is found, while the effect of $\dot{\eta}\neq0$ (magenta diamonds) only appears at large scales. The `back-scaled' case (orange triangles) shows a combination of these behaviours.

The CMC-MD gauge results in the right panel of Figure~\ref{fig:ICs-no-corrections} showing the effect of $\dot{\eta}\neq0$ (green squares) and that of no $\sigma_8$ correction (red stars) are for the $u^i$ velocity obtained via the finite-difference method using the $\delta^p$ power spectrum. Here, we compare the power spectra of $\partial_iu^i_{\rm C}$ (symbols labelled with `(CMC)' in the legend) with those of $\partial_iu^i_{\rm N}$ (symbols labelled with `(N)'), as we expect the two to be equal to each other since $u^i_{\rm C}=u^i_{\rm N}$ (see  Appendix~\ref{appendix:mapping}). The default case, the relative differences of all the other cases with respect to which are shown in the lower subpanel, is `2LPTic (N)', which represents the velocity field generated using the {\sc 2lpt}ic code without back-scaling. The deviations from the default case can be understood as the effects of $\dot{\eta}\neq0$ or not applying the $\sigma_8$ correction. For example, for $\partial_iu^i_{\rm N}$ (cyan circles), not correcting for $\sigma_8$ causes a constant shift; the same constant shift appears for $\partial_iu^i_{\rm C}$ (red stars) at $k\gtrsim0.01h$Mpc$^{-1}$, but on even larger scales the deviation gets larger. On the other hand, not enforcing $\dot{\eta}=0$ while applying the $\sigma_8$ correction (purple diamonds and green squares) leads to good agreement with the default case on small scales, whilst deviations still remain on large scales $k\lesssim0.02h$Mpc$^{-1}$.

\section{Conclusion}\label{sec:conclusion}

This is the second in a series of papers on {\sc gramses}, in which we have addressed the generation of ICs data for this code. We have revised the standard method where the calculations of particles' positions and velocities are both done using the displacement vector, highlighting its limitations when it comes to models beyond $\Lambda$CDM or gauges other than synchronous gauge overdensity and Newtonian gauge velocity, $\{\delta^{\rm S},u^i_{\rm N}\}$. In order to overcome these issues, we have proposed a finite difference calculation for the particles' velocity such that no explicit parameterisation of the growth factor (and growth rate) is required.

In this approach, not only do we need the matter power spectrum $P_{\delta\delta}$ at the initial redshift $z_{\rm ini}$, but it is also required at two neighbouring redshifts (one slightly higher and the other slightly lower). Then, an ICs code is applied to realise these three power spectra (using the same random number seed) to calculate the particle positions at the three redshifts, the central one of which is used as the real particle positions at $z_{\rm ini}$, while the velocities of the particles are calculated by finite-differencing their displacements in the two neighbouring snapshots (this is the simplest way to do finite difference, and more accurate ways are also possible although they generally require more snapshots to be generated). In this way, the basic assumption of a scale-independent linear growth rate of the usual ICs codes (for Newtonian simulations) is avoided, and the method can be applied to generate the initial conditions for any model -- as long as the pair $\{\delta,u^i\}$ used obey the formal continuity equation Eq.~\eqref{eq:LPT-continuity} -- since all the information needed is drawn from the three input matter power spectra. For illustration, we have implemented this finite difference method of the velocity in a modified 2{\sc lptic} code, and the matter power spectrum for the relevant gauge is calculated using a modified version of the Boltzmann code {\sc camb}. The implementation is straightforward, involving minimal modifications to the default 2{\sc lpt}ic code, and we expect this to be true for other standard $N$-body ICs codes.

We have discussed additional steps to remove the radiation effects that might propagate from the power spectra from the Boltzmann code to the $N$-body initial condition generated by this approach, as $N$-body simulations concern only dark matter. These are related to the dynamics of the synchronous-gauge variable $\eta$ and can become non-negligible at large scales. In order to compare with the default 2{\sc lpt}ic code, we have also discussed how to correct the $\sigma_8$ value at the two neighbouring redshifts $z_\pm$. Then, by measuring the matter and velocity divergence power spectra we have shown that the finite difference method allows to recover the ICs as generated by the default 2{\sc lptic} code with sub-percent accuracy at all probed scales and $\Delta z$-values, independently of the spatial resolution. For the case of ICs for the CMC-MD gauge, we have compared against the linear-theory predictions for $P_{\delta\delta}$ and $P_{\theta\theta}$, finding also good agreement. Since this method calculates the upper-index velocity, $u^i$, we have also described an additional step carried out in {\sc gramses} itself before the simulation starts which allows to calculate $u_i$, since this is the variable that is actually used in the $3+1$ form of the geodesic equation for particles. The ICs generated this way have been used to run the $\Lambda$CDM GR cosmological simulations with {\sc gramses} discussed in the first paper of this series, Ref.~\cite{Barrera-Hinojosa:2019mzo}.

\acknowledgments

The authors are grateful to the anonymous referee for their feedback and comments, which helped to clarify a number of points from the original manuscript. 
CB-H is supported by the Chilean National Commission for Scientific and Technological Research through grant CONICYT/Becas-Chile (No.~72180214). BL is supported by the European Research Council (Grant No.~ERC-StG-716532-PUNCA) and STFC Consolidated Grant (Nos. ST/I00162X/1, ST/P000541/1). This work used the DiRAC@Durham facility managed by the Institute for Computational Cosmology on behalf of the STFC DiRAC HPC Facility (\url{www.dirac.ac.uk}). The equipment was funded by BEIS via STFC capital grants ST/K00042X/1, ST/P002293/1, ST/R002371/1 and ST/S002502/1, Durham University and STFC operation grant ST/R000832/1. DiRAC is part of the UK National e-Infrastructure.

\appendix

\section{Mapping of linear equations from the CMC-MD gauge}\label{appendix:mapping}
In this Appendix we include further details on the mapping between the linearised version of {\sc gramses} equations and their standard synchronous gauge and Newtonian gauge counterparts. {For simplicity, in this discussion we focus on scalar perturbations only, so that the gauge transformation can be written as $\xi_\mu=(\xi_0,\partial_i\xi)$, but the results can be extended to include vector modes straightforwardly.}

\subsection{Field equations}

Let us show the correspondence between the field equations in the synchronous and CMC-MD gauges at linear level. We shall ignore the prime notation used for the gauge-transformed variables in Section~\ref{sec:gauge-transform} and instead use $\Phi, \beta_i, \Psi$ and $h_{ij}$ to denote the metric perturbations in the CMC-MD gauge. For the synchronous gauge metric Eq.~\eqref{eq:sync_metric}, we use the variables $h$, $h_{ij}^\parallel$, $\eta$ and $\mu$ following exactly the convention of Ref.~\cite{MB95}. Using that, at linear order, the MD gauge condition Eq.~\eqref{eq:MD-condition} reduces to $\partial^i h_{ij} = 0$, {from Eq.~\eqref{eq:h_ij-gauge-transform} we find} the following condition for the spatial transformation $\xi$
\begin{equation}
    \label{eq:cmc_xi}2a^{-2}\xi = h+6\eta\,.
\end{equation}
Likewise, because the CMC gauge is defined by a condition over $K$, it is useful to derive the explicit gauge transformation for this quantity. The extrinsic curvature at the linear level is given by
\begin{align}
K_{ij}=-H(1-\Phi)\gamma_{ij}+\dot{\Psi}\gamma_{ij}-\frac{a^2}{2}\dot{h}_{ij}+\frac{1}{2}(\partial_i\beta_j+\partial_j\beta_i)\,,\label{eq:Kij-calc}
\end{align}
and its trace given by $K=-3H(1-\Phi)+3\dot{\Psi}+a^{-2}\delta^{ij}\partial_i\beta_j$. Then, using the gauge transformations Eqs.~\eqref{eq:alpha-gauge-transform}-\eqref{eq:cmc_psi} on the right-hand side of the latter, we find that $K$ transforms as
\begin{equation}\label{eq:K-gauge-transform}
K=K'+3\dot{H}\xi_0+\gamma^{ij}\partial_i\partial_j\xi_{0}\,.
\end{equation}
As expected, Eq.~(\ref{eq:K-gauge-transform}) is independent of $\xi^i$ and can be used to connect the time coordinate defined by the CMC foliation with that in any other gauge regardless of the choice of spatial coordinates.  For the case of synchronous gauge, using that $K_{\rm S}=-3H-\dot{h}/2$ and Eq.~\eqref{eq:CMC}, we find 
\begin{equation}
\dot{h}=-6\dot{H}\xi_0-2\gamma^{ij}\partial_i\partial_j\xi_0 \,,\label{eq:cmc_gauge_condition}
\end{equation}
where $\gamma^{ij}\partial_{i}\partial_{j}=a^{-2}\delta^{ij}\partial_{i}\partial_{j}\equiv a^{-2}\partial^2$.

Let us now consider the Hamiltonian constraint and Eq.~\eqref{eq:cmc-elliptic}. These are given at the linear order by
\begin{eqnarray}
    \label{eq:cmc_psi_eqn}\partial^2\Psi &=& 4\pi Ga^2\delta\rho,\\
    \label{eq:cmc_alpha_eqn}\partial^2\Phi + 3\dot{H}a^2\Phi &=& 4\pi Ga^2\delta\rho + 12\pi Ga^2\delta P,
\end{eqnarray}
where $\delta\rho=\bar{\rho}\delta^{\rm C}$ with $\delta^{\rm C}$ is the density contrast in the CMC-MD gauge. Using the gauge transformations Eq.~\eqref{eq:cmc_psi} and Eq.~\eqref{eq:delta-gauge-transf}, Eq.~\eqref{eq:cmc_psi_eqn} can be rewritten in terms of the synchronous gauge and gauge transformation variables as
\begin{equation}\label{eq:cmc_to_syn}
    -\frac{1}{6}\partial^2h - H\partial^2\xi_0 + \frac{1}{3}a^{-2}\partial^4\xi = 4\pi Ga^2\bar{\rho}\delta^{\rm S} - 12\pi Ga^2H(\bar{\rho}+\bar{P})\xi_0\,.
\end{equation}
Applying the gauge relations Eq.~(\ref{eq:cmc_xi})
and Eq.~(\ref{eq:cmc_gauge_condition}), the left-hand side of this equation becomes
\begin{equation}\label{eq:cmc_to_syn-2}
    -H\partial^2\xi_0 = \frac{1}{2}a^2H\dot{h}+3H\dot{H}a^2\xi_0 = \frac{1}{2}a^2H\dot{h}-12\pi Ga^2(\bar{\rho}+\bar{P})H\xi_0\,,
\end{equation}
where in the second equality we have used the background relation
\begin{equation}\label{eq:Hdot-background}
     \dot{H} = -4\pi G(\bar{\rho}+\bar{P})\,.
\end{equation}
Therefore, Eq.~(\ref{eq:cmc_to_syn-2}) can be simplified as
\begin{equation}
    \partial^2\eta + \frac{1}{2}a^2H\dot{h} = 4\pi Ga^2\bar{\rho}\delta^{\rm S}\,,
\end{equation}
which is the $(00)$ Einstein equation in synchronous gauge~\cite{MB95}.

Next, let us consider how Eq.~\eqref{eq:cmc_alpha_eqn} transforms. Using the gauge transformations Eq.~\eqref{eq:alpha-gauge-transform} and Eq.~\eqref{eq:delta-gauge-transf}, as well as the gauge transformation for the pressure perturbation $\delta P^{\rm C}=\delta P^{\rm S}+\dot{\bar{P}}\xi_0$, Eq.~\eqref{eq:cmc_alpha_eqn} can be rewritten as
\begin{equation}
    \partial^2\dot{\xi}_0 + 3\dot{H}a^2\dot{\xi}_0 = 4\pi Ga^2\delta\rho^{\rm S} - 12\pi Ga^2H(\bar{\rho}+\bar{P})\xi_0 + 12\pi Ga^2\delta P^{\rm S} + 12\pi Ga^2\dot{\bar{P}}\xi_0\,.
\end{equation}
In order to eliminate $\xi_0$ and $\dot{\xi}_0$ from the left-hand side of this equation, we take the time derivative of the gauge relation Eq.~(\ref{eq:cmc_gauge_condition}), to get
\begin{equation}\label{eq:h_ddot-temp}
    -\frac{1}{2}\ddot{h} - H\dot{h} = 4\pi G\left(\delta\rho^{\rm S}+3\delta P^{\rm S}\right) - 12\pi GH(\bar{\rho}+\bar{P})\xi_0 + 3\ddot{H}\xi_0 + 6H\dot{H}\xi_0 + 12\pi G\dot{\bar{P}}\xi_0\,,
\end{equation}
and taking the time derivative of Eq.~\eqref{eq:Hdot-background} allows to get rid of $\ddot{H}$ and all terms proportional to $\xi_0$ in the right-hand side of Eq.~\eqref{eq:h_ddot-temp}, leaving
\begin{equation}
    -\frac{1}{2}\ddot{h} - H\dot{h} = 4\pi G\left(\delta\rho^{\rm S}+3\delta P^{\rm S}\right)\,,
\end{equation}
which is equivalent to the linear combination of Einstein equations $2\times(00)+(ii)$ in synchronous gauge~\cite{MB95} (where $(ii)$ denotes the trace of the $(ij)$ components of the Einstein equation).

Next, let us consider the momentum constraint, which gives the longitudinal part of the conformal curvature tensor $\bar{A}_{ij}$. This is given at linear order by~\cite{Barrera-Hinojosa:2019mzo}
\begin{align}
    \label{eq:cmc_w_eq1}\left(\bar{\Delta}_LW\right)_i=\ \partial^2W_i + \frac{1}{3}\partial_i\delta^{kl}\partial_kW_l = 8\pi Ga^3\left(\bar{\rho}+\bar{P}\right)u_i^{\rm C}\,,
\end{align}
with
\begin{align}
    \label{eq:cmc_w_eq2}\bar{A}_{ij}=\left(\bar{L}W\right)_{ij}\equiv\partial_{i}W_{j}+\partial_{j}W_{i}-\frac{2}{3}\delta_{ij}\partial_kW^k \,.
\end{align}
We will use Eq.~(\ref{eq:cmc_w_eq2}) to solve $W_i$ first, and then substitute into Eq.~(\ref{eq:cmc_w_eq1}) to get an equation that contains the peculiar velocity. The right-hand side of Eq.~(\ref{eq:cmc_w_eq2}) can be written as
\begin{eqnarray}
    \left(\bar{L}W\right)_{ij} &=& 2\left(\partial_i\partial_j-\frac{1}{3}\delta_{ij}\partial^2\right)W\,,\label{eq:x1}
\end{eqnarray}
where, given that we specialise to the scalar mode only, we have introduced the variable $W$ such that $W_i=\partial_iW$. Using the fact that at the linear level $\bar{A}_{ij}=aA_{ij}$, from the traceless part of (\ref{eq:Kij-calc}) we find
\begin{eqnarray}
    \bar{A}_{ij} &=& -\frac{1}{2}a^3\dot{h}_{ij} + \frac{1}{2}a\left(\partial_i\beta_j+\partial_j\beta_i\right) - \frac{1}{3}a\delta^{kl}\partial_k\beta_l\delta_{ij}\,,
\end{eqnarray}
and, using Eq.~(\ref{eq:beta-gauge-transform}) to get rid of $\beta_i$, as well as Eq.~(\ref{eq:h_ij-gauge-transform}) to get rid of $\dot{h}_{ij}$, we find
\begin{align}
       \bar{A}_{ij}= -a\left(\partial_i\partial_j-\frac{1}{3}\delta_{ij}\partial^2\right)\xi_0 - \frac{1}{2}a^3\left(\partial_i\partial_j-\frac{1}{3}\delta_{ij}\partial^2\right)\dot{\mu}\,,\label{eq:x2}
\end{align}
where $\mu$ is a synchronous gauge scalar perturbation variable whose relation with $h,\eta$ is given below. Combining Eq.~(\ref{eq:x1}) and Eq.~(\ref{eq:x2}) gives the result
\begin{equation}\label{eq:cmc_w}
W = -\frac{1}{2}a\xi_0 - \frac{1}{4}a^3\dot{\mu}\,.
\end{equation}
Using the gauge transformation for $u_i$, Eq.~\eqref{eq:ulowi-gauge-transf}, in the right-hand side of Eq.~\eqref{eq:cmc_w_eq1}, then applying $\partial^i$ on both sides and substituting the result Eq.~\eqref{eq:cmc_w}, we get
\begin{eqnarray}
-\frac{2}{3}a\partial^2\partial^2\xi_0 - \frac{1}{3}a^3\partial^4\dot{\mu}
= 8\pi Ga^3\left(\bar{\rho}+\bar{P}\right)\left(\partial^iu_i^{\rm S}-\partial^2\xi_0\right)\,.
\end{eqnarray}
Now, using Eq.~\eqref{eq:cmc_gauge_condition} to eliminate $\partial^2\partial^2\xi_0$ from the left-hand side of this equation, as well as the synchronous gauge relation $\partial^2\mu=h+6\eta$, we have 
\begin{align}
\frac{1}{3}a^3\partial^2\dot{h} + 2\dot{H}a^3\partial^2\xi_0 - \frac{1}{3}a^3\partial^2\left(\dot{h}+6\dot{\eta}\right)
=& 8\pi Ga^3\left(\bar{\rho}+\bar{P}\right)\left(\partial^iu_i^{\rm S}-\partial^2\xi_0\right)\,.
\end{align}
Finally, by using the first and second Friedmann equations, we can eliminate the terms proportional to $\partial^2\xi_0$ from the above equation, which reduces to
\begin{equation}
    -2\partial^2\dot{\eta} = 8\pi G(\bar{\rho}+\bar{P})\theta^{\rm S}\,,
\end{equation}
where $\theta^{\rm S}=\partial^iu_i^{\rm S}$. This corresponds to the $(0i)$ Einstein equation in synchronous gauge~\cite{MB95}.

\subsection{Equations of motion}

Let us provide some complementary derivation about the relation $u^i_{\rm C}=u^i_{\rm N}$ mentioned in Section~\ref{sec:gauge-transform}. For this, we consider the geodesic equations for non-relativistic particles, which at the linear level are given by Eqs.~\eqref{eq:geodesic-1} and \eqref{eq:geodesic-2}. We next show that in linear perturbation regime, the equation for $u^i_{\rm C}$ is identical to that of $u^i_{\rm N}$ in the Newtonian gauge. 

Inverting the relation in Eq.~\eqref{eq:uupperi-uloweri} for $u^i$ and taking one time derivative, we have 
\begin{eqnarray}
\dot{u}^i_{\rm C} = \frac{\rm d}{{\rm d}t}\left(\gamma^{ij}u^{\rm C}_j-\beta^i\right) = -2H\gamma^{ij}u^{\rm C}_j - \gamma^{ij}\partial_j\Phi- \dot{\beta}^i\,,
\end{eqnarray}
where in the second step we have used Eq.~\eqref{eq:geodesic-1} to get rid of $\dot{u}^{\rm C}_j$. Now, using Eq.~\eqref{eq:uupperi-uloweri} and  $\beta^i= a^{-2}\delta^{ij}\beta_j$, we obtain
\begin{align}
\dot{u}^i_{\rm C}= -2Hu^i_{\rm C} - \gamma^{ij}\partial_j\Phi - \gamma^{ij}\dot{\beta}_j\,.\label{eq:temp5}
\end{align}
To make a connection with the Newtonian gauge we consider the metric Eq.~\eqref{eq:newt_metric}. From the gauge transformations Eq.~\eqref{eq:alpha-gauge-transform}-\eqref{eq:h_ij-gauge-transform} we find the following relations between the CMC-MD and Newtonian gauge metric perturbations
\begin{eqnarray}
\label{eq:tmp1a}\Phi &=& \psi + \dot{\xi}_0,\\
\label{eq:tmp2a}\beta_i &=& -\dot{\xi}_i - \partial_i\xi_0 + 2H\xi_i,\\
\Psi &=& \phi - H\xi_0 + \frac{1}{3}\gamma^{ij}\partial_j\xi_{i},\\
h_{ij} &=& -a^{-2}\left(\partial_j\xi_{i}+\partial_i\xi_{j}\right) + \frac{2}{3}\gamma^{kl}\partial_l\xi_{k}\delta_{ij}\,.\label{eq:h_i_j-Newt-transf}
\end{eqnarray}
By using the MD gauge condition $\partial^ih_{ij}=0$ in Eq.~\eqref{eq:h_i_j-Newt-transf}, we find $\xi_i=0$ in this case. Then, after applying the gauge transformations for $\Phi$ and $\beta_i$, Eqs.~\eqref{eq:tmp1a} and \eqref{eq:tmp2a}, the equation of motion Eq.~\eqref{eq:temp5} becomes
\begin{equation}
    \dot{u}^i_{\rm C} + 2Hu^i_{\rm C} + \gamma^{ij}\partial_j\psi = 0\,,
\end{equation}
which is identical to the Newtonian gauge counterpart,
$\dot{u}^i_{\rm N} + 2Hu^i_{\rm N} + \gamma^{ij}\partial_j\psi = 0$. This equation-level equivalence is not surprising given that $u^i_{\rm C}=u^i_{\rm N}$, which is a consequence of $\xi_i=0$.

To make connection with synchronous gauge, we combine the geodesic equations to eliminate $u_i$ and get a second order differential equation for $x^i$,
\begin{equation}\label{eq:cmc_geodesic}
    \frac{{\rm d}^2x^i}{{\rm d}t^2} + 2H\frac{{\rm d}x^i}{{\rm d}t} + a^{-2}\delta^{ij}\partial_j\Phi + a^{-2}\delta^{ij}\dot{\beta}_j = 0\,,
\end{equation}
which is for the CMC-MD gauge. We can rewrite this equation in terms of the gauge transformation and synchronous gauge variables, using Eqs.~\eqref{eq:alpha-gauge-transform} and \eqref{eq:beta-gauge-transform}, to find
\begin{equation}
    \partial_i\Phi+\dot{\beta}_i = -\partial_i\ddot{\xi} + 2\dot{H}\partial_i\xi + 2H\partial_i\dot{\xi}\,.
\end{equation}
The particle coordinate in the synchronous gauge, denoted as $x^i_{\rm S}$, is related to the CMC-MD gauge coordinate $x^i$ as $x^i=x^i_{\rm S}+a^{-2}\delta^{ij}\partial_j\xi$. Taking the time derivatives of the latter and substituting the resulting expressions into Eq.~(\ref{eq:cmc_geodesic}) gives
\begin{equation}
    \frac{{\rm d}^2x^i_{\rm S}}{{\rm d}t^2} + 2H\frac{{\rm d}x^i_{\rm S}}{{\rm d}t} = 0\,,
\end{equation}
as we expect to find in the synchronous gauge.

\bibliographystyle{JHEP}

\providecommand{\href}[2]{#2}\begingroup\raggedright\endgroup
\end{document}